\documentclass[aps,prb,showpacs,floatfix,twocolumn,superscriptaddress,citeautoscript]{revtex4-1}
\usepackage{graphicx}
\usepackage{amsmath}
\usepackage{amssymb}
\usepackage{bm}
\usepackage{dcolumn}
\usepackage{subfigure}
\usepackage{units}
\usepackage{hyperref}
\usepackage[usenames]{color}
\hypersetup{pdfborder=0 0 0,colorlinks=true,citecolor=blue,linkcolor=blue}
\input epsf
\newcommand{\BN}{\textit{h}-BN}

\begin{document}

\title{Field effect doping of graphene in metal$|$dielectric$|$graphene heterostructures: a model based upon  first-principles calculations}
\author{Menno Bokdam}
\affiliation{Faculty of Science and Technology and MESA$^{+}$ Institute for Nanotechnology, University of Twente, P.O. Box 217, 7500 AE Enschede, The Netherlands}
\author{Petr A. Khomyakov}
\affiliation{IBM Research GmbH, Z$\ddot u$rich Research Laboratory, S$\ddot a$umerstrasse 4, 8803 R$\ddot u$schlikon, Switzerland}
\author{Geert Brocks}
\affiliation{Faculty of Science and Technology and MESA$^{+}$ Institute for Nanotechnology, University of Twente, P.O. Box 217, 7500 AE Enschede, The Netherlands}
\author{Paul J. Kelly}
\affiliation{Faculty of Science and Technology and MESA$^{+}$ Institute for Nanotechnology, University of Twente, P.O. Box 217, 7500 AE Enschede, The Netherlands}

\begin{abstract}
We study how the Fermi energy of a graphene monolayer separated from a conducting substrate by a dielectric spacer depends on the properties of the substrate and on an applied voltage. An analytical model is developed that describes the Fermi level shift as a function of the gate voltage, of the substrate work function, and of the type and thickness of the dielectric spacer. The parameters of this model, that should describe the effect of gate electrodes in field-effect devices, can be obtained from density functional theory (DFT) calculations on single layers or interfaces. The doping of graphene in metal$|$dielectric$|$graphene structures is found to be determined not only by the difference in work function between the metal and graphene and the dielectric properties of the spacer but potential steps that result from details of the microscopic bonding at the interfaces also play an important role. The doping levels predicted by the model agree very well with the results obtained from first-principles DFT calculations on metal$|$dielectric$|$graphene structures with the metals Al, Co, Ni, Cu, Pd, Ag, Pt or Au, and a \BN{} or vacuum dielectric spacer.     
\end{abstract}
\date{\today}
\maketitle

\section{Introduction}\label{sec:intro}

The successful extraction of graphene, single monolayers of graphite, has led to the discovery of new physical phenomena that result from the unique electronic structure of this two-dimensional system.\cite{Novoselov:sc04} Many of the remarkable properties of graphene have been explored using field effect devices \cite{Novoselov:nat05,Zhang:nat05} in which a graphene flake is placed on top of an oxidized silicon wafer and contacted with metal (source and drain) electrodes. 

Direct contact between graphene and a metal electrode generally results in a transfer of electrons between the two materials. Depending upon the metal species used, the contact causes graphene to be doped $n$- or $p$-type in equilibrium. \cite{Giovannetti:prl08,Khomyakov:prb09}  
Metal-contact doping also induces a lateral space charge and corresponding band bending in the graphene adjacent to, but not covered by, the metal contact. The two-dimensional character of graphene and the low density of states (Dos) around its Fermi level then result in an extended space charge region.\cite{Khomyakov:prb10} The space charge distribution is inhomogeneous, which can lead to $p$-$p^{\prime}$, $n$-$n^{\prime}$, $p$-$n$ or, with gating, to even more complex profiles such as $p$-$n$-$p^{\prime}$ within the space charge region.\cite{Lee:natn08,Khomyakov:prb10,Wu:nanol12} 

In this paper we focus on electrostatic doping, where application of a voltage difference between graphene and a back gate changes the position of the Fermi energy with respect to the Dirac point \cite{Novoselov:nat05,Huard:prb08,Zhang:natp08,Zhang:natp09,Yu:nanol09,Britnell:sci12} and a nonlinear shift of the graphene Fermi energy as a function of the gate voltage has been observed in scanning tunneling spectroscopy (STS) experiments\cite{Zhang:natp08} and in work function measurements.\cite{Yu:nanol09} In experiments such as these, where a SiO$_2$ dielectric was used with a doped Si gate, it is believed that intrinsic inhomogeneities in SiO$_2$ as well as surface roughness are reasons that much lower charge carrier mobilities are observed than for freely suspended graphene.\cite{Bolotin:ssc08} 

In this respect, a much better gate dielectric for graphene based electronics is hexagonal boron-nitride, \BN{}.\cite{Giovannetti:prb07,Karpan:prb11,Bokdam:nanol11,Dean:natn10,Xue:natm11,Decker:nanol11,Britnell:sci12} Graphene adsorbed on \BN{} has been shown \cite{Dean:natn10,Xue:natm11,Decker:nanol11} to be much flatter than on SiO$_2$ and to have highly mobile charge carriers.\cite{Dean:natn10} This favorable situation can be related to the fact that \BN{} has a layered crystal structure similar to that of graphite; it consists of atoms that are strongly bonded in plane in a honeycomb lattice, whereas the binding between these planes is much weaker. But where graphene is a semi-metal, \BN{} is an insulator. Graphene binds weakly to \BN{} with an interaction similar to that between \BN{} monolayers, or between the graphene layers in graphite.\cite{Giovannetti:prb07} All of this means that adsorbing graphene onto \BN{} perturbs its unique electronic structure minimally.

Like (multilayer) graphene, layers of \BN{} can be prepared by micromechanical cleavage,\cite{Novoselov:pnas05} and can be transferred to a metal substrate to form a gate electrode.\cite{Lee:apl11,Britnell:nanol12} Alternatively, single layers of \BN{} have been grown directly on the (111) surface of transition metals such as Cu, Pt, Rh, Pd and Ni.\cite{Greber:ss09,Preobrajenski:ss05,Cavar:ss08} Graphene$|$\BN{} heterostructures have been grown on Ni(111)\cite{Oshima:ssc00} and more recently on Ru(0001).\cite{Bjelkevig:jpcm10} Very recently, structures of alternating \BN{} and graphene layers have been constructed and shown to operate as a field-effect tunneling transistor.\cite{Britnell:sci12}

Obviously the properties of the gate electrode play an important role in determining the behavior of graphene field effect devices. In this paper we develop an analytical model for the electrostatic doping of graphene as a function of the electric field generated by a gate bias and test the model against results obtained using first-principles density functional theory (DFT) calculations on metal gate$|$dielectric$|$graphene (M$|$D$|$Gr) stacks in the presence of an external electric field across the stack. The results of calculations for the particular case of M = Cu were recently briefly reported.\cite{Bokdam:nanol11}

The electrostatic model for the M$|$D$|$Gr structure is based upon a parallel plate capacitor geometry. The input parameters to the model are the work function of the metal gate, the dielectric constant and thickness of the gate dielectric, and the potential steps formed at the interfaces between the metal and the dielectric, and between the dielectric and the graphene sheet. Such potential steps originate from interactions between the two materials at an interface, and they can make a non-negligible contribution to the potential profile.\cite{Giovannetti:prl08,Khomyakov:prb09,Rusu:jpcc09,Rusu:prb10,Bokdam:nanol11} All the parameters are accessible to DFT calculations on single layers or interfaces.

In the DFT calculations on the full M$|$D$|$Gr structures, we use primarily \BN{} as dielectric material and study the series of metals Al, Co, Ni, Cu, Pd, Ag, Pt, and Au, whose work functions span a considerable range, from 4.2 to 6.0 eV. \BN{} is physisorbed on Al, Cu, Ag, Pt, and Au and chemisorbed on Co, Ni, and Pd allowing us to study the effect of the bonding between metal and dielectric on the electrostatic doping of graphene. To assess the influence of the dielectric, we examine the effect of using a vacuum ``dielectric'' instead of \BN{}. 

The paper is organized as follows. In the next section we formulate the analytical model for electrostatic doping of graphene. We describe the DFT calculations and the (electronic) structures of the metal$|$\BN{}$|$graphene stacks in Sec.~\ref{sec:compdetails}. The results of the calculations are presented in Sec.~\ref{sec:results} and compared to the predictions of the model. Finally, Sec.~\ref{sec:discussion} presents a summary plus some conclusions.

\section{Electrostatic doping model}\label{sec:model}

\begin{figure}
\includegraphics[scale=0.35]{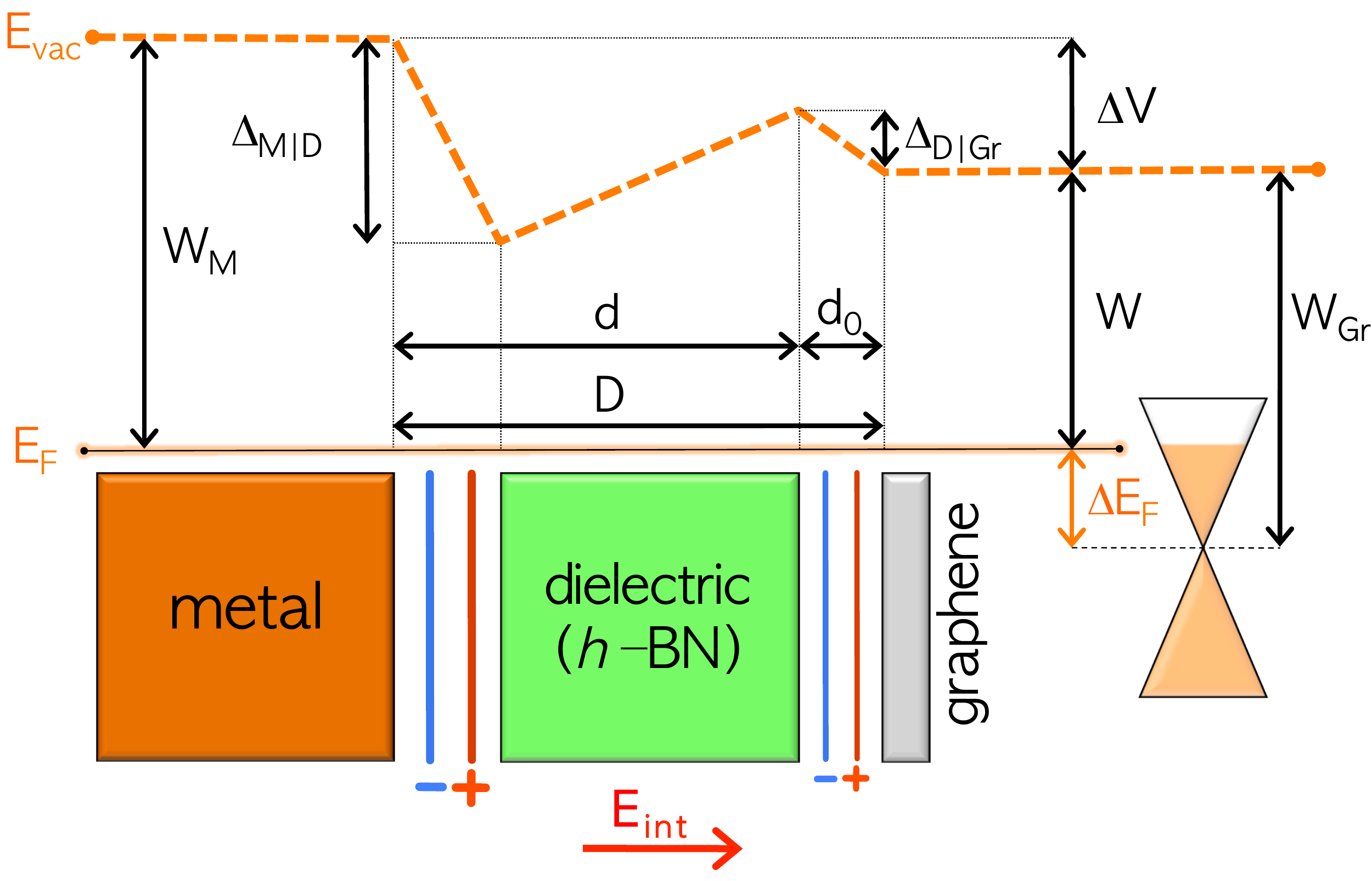}
\caption{(color online) Schematic drawing of the metal$|$dielectric$|$graphene (M$|$D$|$Gr) structure. The electrostatic potential profile in the structure is indicated by the wide dashed (orange) line at the top. $\Delta V$ is the shift of the work function of the structure with respect to that of the clean metal, $W_\mathrm{M}$; $\Delta_\mathrm{M|D},\Delta_\mathrm{D|Gr}$ are the potentials steps at the interfaces; $\Delta E_\mathrm{F}$ is the Fermi level shift in graphene caused by the electrostatic doping.}
\label{fig:potential}
\end{figure}

We model the M$|$D$|$Gr structure as a parallel plate capacitor. Such a model has been used successfully to explain the Fermi level shift in graphene directly absorbed on a metal substrate.\cite{Giovannetti:prl08,Khomyakov:prb09} Inserting a dielectric is expected to modify the potential profile as shown schematically in Fig.~\ref{fig:potential}. The shift of the Fermi level, $\Delta E_{\rm F}$, with respect to the charge neutrality point in graphene, i.e. the conical points of the band structure, is given by
\begin{equation}
\Delta E_{\rm F} = W - W_{\rm Gr}, \label{eq1}
\end{equation}
with $W_{\rm Gr}$ and $W$ the work functions of graphene in vacuum, and of the complete M$|$D$|$Gr stack, respectively. Note that a negative value of $\Delta E_{\rm F}$ represents electron doping, a positive value hole doping. $W$ can be written as
\begin{equation}
 W = W_{\rm M} - \Delta V, \label{eq2}
\end{equation}
with $W_{\rm M}$ the work function of the clean metal substrate, and $\Delta V$ the potential difference across the stack. The latter will in general be a function of the dielectric thickness $d$. 

We model the potential difference as
\begin{equation}
\Delta V = -e E_{\rm int} d + \Delta_{\rm M|D}+ \Delta_{\rm D|Gr}, \label{eq3}
\end{equation}
where $E_{\rm int}$ is the electric field inside the dielectric, $-e$ is the charge of an electron, and $\Delta_{\rm M|D}$ and $\Delta_{\rm D|Gr}$ are potential steps at the metal$|$dielectric and the dielectric$|$graphene interfaces, respectively. Interface bonding between two materials usually results in the formation of a dipole layer that is accompanied by a potential step\cite{Rusu:jpcc09,Rusu:prb10,Rusu:prb06} localized at the interface. Interface dipoles are intrinsic properties of interfaces and even if the interaction at an interface is as weak as in the case of physisorption, they can be sizable. We will assume that interface potential steps are independent of applied electric fields for field magnitudes of practical interest and test this assumption with DFT calculations.

In the absence of a dielectric spacer, an externally applied gate voltage gives rise to an electric field $E_{\rm ext}$. In the following, it will be convenient to use $E_{\rm ext}$ as an independent variable. 
The internal field, Eq.~(\ref{eq3}), is obtained from the standard electrostatic boundary condition
\begin{equation}
\epsilon_{0}\left( E_{\rm ext} - \kappa E_{\rm int} \right) = \sigma, \label{eq4}
\end{equation}
with $\kappa$ the relative dielectric constant of the dielectric layer, and $\sigma$ the surface charge density of the graphene sheet. 

The latter can be related to the Fermi level shift $\Delta E_{\rm F}$ by noting that charge in graphene is introduced by occupying states starting from the charge neutrality point, $\sigma = e\int_0^{\Delta E_{\rm F}} D(E)dE$, where the charge neutrality point corresponds to the conical points in the graphene band structure. The density of states near the conical points is described well by the linear function, $D(E)=|E|\, D_0/A$, with $D_0=0.102$/(eV$^2$ unit cell), and $A=5.18$ \AA$^2$ as the area of a graphene unit cell, so that   
\begin{equation}
\sigma = {\rm sign}(\Delta E_{\rm F}) \frac{eD_{0}}{2A} (\Delta E_{\rm F})^{2}. \label{eq5}
\end{equation}
The value of $D_{\rm 0}$ is sensitive to the $\mathbf{k}$-point grid used for the Brillouin zone sampling, and it depends on the substrate used; technical details are discussed in the Appendix.\cite{fn1}

Equations (\ref{eq1})-(\ref{eq5}) can be combined to give a closed expression for $\Delta E_{\rm F}$ in terms of the applied external field $E_{\rm ext}$
\begin{equation}
\Delta E_{\rm F}(E_{\rm ext})=\pm \frac{\sqrt{1+2D_0\alpha(d/\kappa)^2\left|e(E_{\rm ext}+E_0)\right|}-1}{D_0\alpha d/\kappa},
\label{fermishiftE}
\end{equation}
where $\alpha= e^2/(\epsilon_0 A) = 34.93$ eV/\AA\ for a $1\times 1$ graphene unit cell. The sign of $\Delta E_{\rm F}$ is determined by the sign of $E_{\rm ext}+E_0$, where 
\begin{equation}
E_0=V_{\rm 0}\frac{\kappa}{de},
\label{E0}
\end{equation}
with
\begin{equation}
V_{\rm 0}=W_{\rm M}-W_{\rm Gr}-\Delta_{\rm M|D}-\Delta_{\rm D|Gr}.
\label{eq:V0}
\end{equation}
$E_0$ is the value of the {\em intrinsic} electric field across the dielectric in the absence of any external electric field, see Fig. \ref{fig:potential}, and it can be substantial when $d$ is sufficiently small, i.e., for thin dielectric spacers. It results from an electron transfer in equilibrium between the metal and graphene across the dielectric spacer that leads to an intrinsic doping of graphene. A compensating external field $E_{\rm ext}=-E_0$ is required to bring the Fermi level to the conical points and make graphene electrically neutral, $\Delta E_{\rm F}=0$. 

\begin{figure}
\includegraphics[scale=0.33]{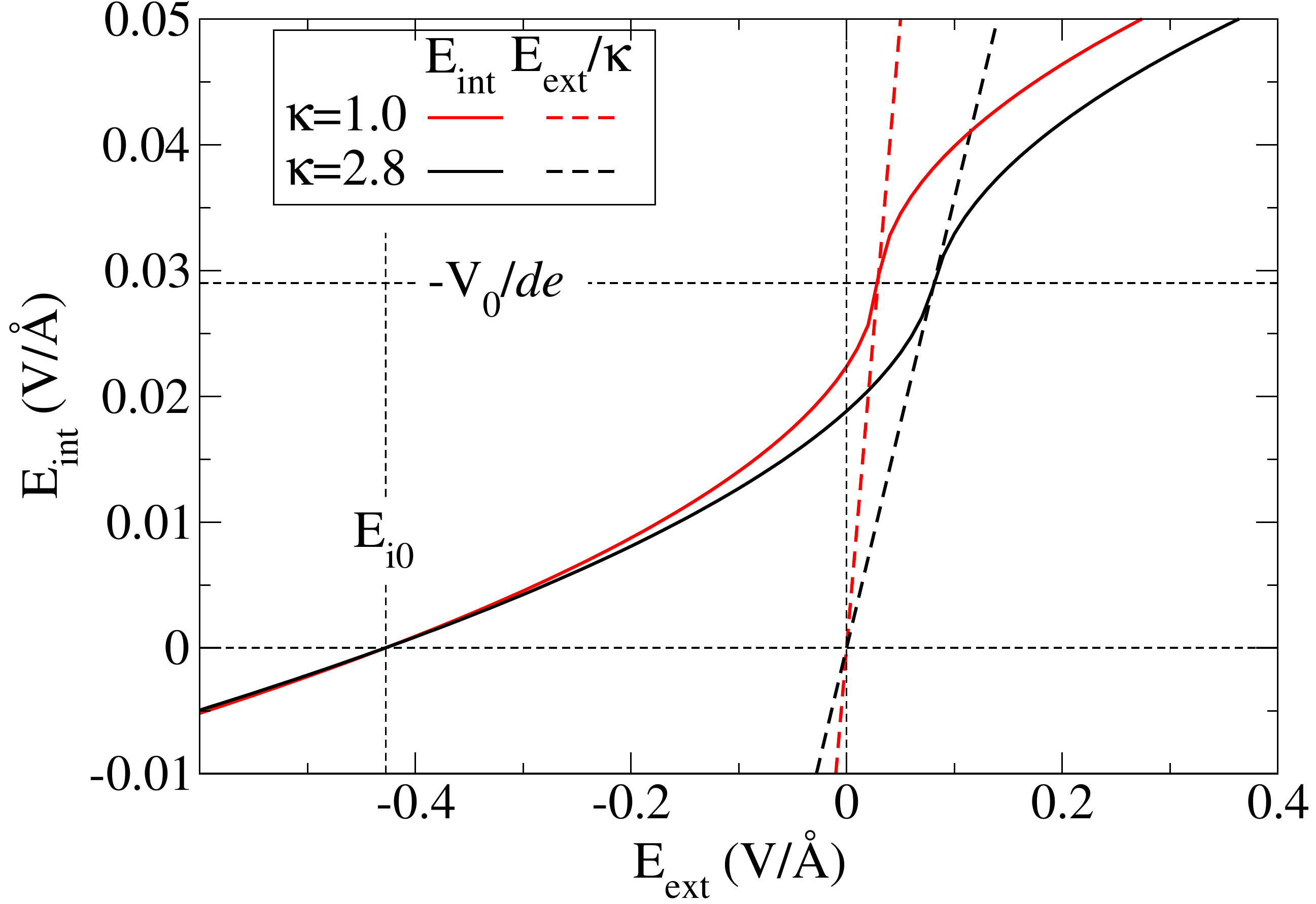}
\caption{(color online) The internal field $E_{\rm int}$ (solid lines) inside the dielectric spacer of Fig. \ref{fig:potential}, as a function of the applied external field $E_\mathrm{ext}$, for two values of $\kappa$. The black line corresponds to a Cu$|$\BN{}$|$Graphene structure with 5 layers of \BN{} and $\kappa=2.8$. The red line would be the response of the same system, but with $\kappa=1.0$ instead of 2.8. The curves cross at $E_\mathrm{ext}=E_{\rm i0}$, see Eq. (\ref{Ei0}), where $E_{\rm int}=0$. At fields $E_\mathrm{ext}=-E_{\rm 0}$, see Eq. (\ref{E0}), the curves cross the lines $E_{\rm ext}/\kappa$ (dashed lines).} 
\label{fig-Eint}
\end{figure}

According to Eqs. (\ref{eq1})-(\ref{eq3}), the Fermi level shift $\Delta E_{\rm F}$ and the internal electric field $E_\mathrm{int}$ are linked by a linear relation.  This means that both these quantities are similar non-linear functions of the external electric field $E_\mathrm{ext}$, Eq. (\ref{fermishiftE}), whose non-linearity is determined by the doping charge on graphene. Figure \ref{fig-Eint} shows $E_\mathrm{int}$ as a function of $E_\mathrm{ext}$ for two different values of the dielectric constant $\kappa$. The points in these curves where the derivative is maximal correspond to undoped graphene, i.e., $\Delta E_{\rm F}=0$ and $E_{\rm ext}=-E_0$. It follows from Eq. (\ref{eq4}) that at these points  the curves cross the lines $E_{\rm ext}/\kappa$, see Fig. \ref{fig-Eint}. From Eq. (\ref{E0}) it follows that the corresponding internal electric field is $E_\mathrm{int} = -V_0/de$.

The $E_{\rm int}$ curves cross at $E_{\rm int}=0$. From Eqs. (\ref{eq1})-(\ref{eq5}) it is easily seen that this occurs when $E_{\rm ext}=E_{\rm i0}$, where
\begin{equation}
E_{\rm i0}=V_{\rm 0}|V_{\rm 0}|\frac{eD_{\rm 0}}{2A\epsilon_0}.
\label{Ei0}
\end{equation}
The corresponding Fermi level shift 
\begin{equation}
\Delta E_{\rm F}(E_{\rm i0}) = V_0,
\label{eq10}
\end{equation}
is independent of $d$ and $\kappa$, i.e., all curves $\Delta E_{\rm F}(E_{\rm ext})$ for different dielectric thicknesses cross at $\Delta E_{\rm F}=V_0$ for $E_{\rm ext}=E_{\rm i0}$. This crossing point could be used to determine $V_{\rm 0}$ from experiment.

Equation (\ref{fermishiftE}) constitutes the model we will use in the remainder of this paper. Assuming that the properties of graphene are known, the model depends on the parameters $W_{\rm M}$, $\Delta_{\rm M|D}$, $\Delta_{\rm D|Gr}$, and $\kappa$. All of these parameters can be obtained from DFT calculations. The thickness of the dielectric spacer $d$ is somewhat ill-defined at the atomic level. As in Ref. \onlinecite{Khomyakov:prb09}, we set $d=D-d_0$, where $D$ is the distance from the center of the top layer of metal atoms to the graphene plane, and $d_{\rm 0}=\rm{2.4}$ \AA.

The model can also be expressed in terms of an applied gate voltage, writing the latter as
\begin{equation}\label{Vg}
V_{\rm g} = -e E_{\rm ext}d/\kappa.
\end{equation}
Equation \eqref{fermishiftE} then becomes
\begin{equation}
\Delta E_{\rm F}=\pm\frac{\sqrt{1 + 2\alpha D_0\, d\left\vert V_{\rm g} - V_0 \right\vert/\kappa}  - 1}{\alpha D_0\, d/\kappa},
\label{modelV} 
\end{equation} 
and the sign of $\Delta E_{\rm F}$ is determined by the sign of $V_{0} - V_{\rm g}$.

\section{Computational details}\label{sec:compdetails}

In this section we briefly describe the most salient features of the first-principles DFT calculations that we use to determine the parameters entering the model defined in the previous section. We calculate electronic ground state energies within the framework of the plane-wave PAW pseudopotential method,\cite{Blochl:prb94b} at the level of the local density approximation (LDA),\cite{Perdew:prb81} as implemented in VASP\cite{Kresse:prb99,Kresse:prb93,Kresse:prb96} and use them to optimize the geometries of M$|$D$|$Gr structures for a number of representative metals. The structures are modeled as periodic slabs separated by a vacuum region $\sim 15$ {\AA} thick. A dipole correction is applied to avoid spurious interactions between periodic images of the slab.\cite{Neugebauer:prb92} A plane wave kinetic energy cutoff is set at 400 eV. 

A $36\times 36$ \textbf{k}-point grid is used to sample the surface Brillouin Zone (BZ) and structural relaxations are performed using a smearing method for the BZ integration,\cite{Methfessel:prb89} with a smearing parameter of 0.2 eV. Total energies and charge densities of optimized structures are then recalculated using the tetrahedron scheme. The special points $\Gamma $, $K$, $K'$ and $M$ of the graphene band structure are explicitly included in the BZ sampling, which is important for an accurate description of the electronic structure around these points. 

The electronic self-consistency criterion is set to $10^{-8}$ eV. Such a strict criterion is required to assure convergence of the charge distribution. We consider only close-packed metal surfaces with three-fold rotational symmetry, i.e., the (111) surface for fcc metals, and the (0001) surface for hcp metals. A slab of six layers of atoms is used to represent the metal surface. The \BN{} layers are adsorbed on one side of the slab to represent the dielectric, and graphene is adsorbed on top of this. To model the effect of the gate potential, a sawtooth-shaped potential is applied across the whole slab.\cite{Resta:prb86} The carbon atoms in graphene, the boron and nitrogen atoms in the {\it h}-BN slab, and the separation between the top two atomic layers of the metal were allowed to relax during the geometry optimization until the total energies were converged to within $10^{-7}$ eV. 

The LDA is found to give a reasonable description of the weak interaction between \BN{} layers and between a \BN{} layer and graphene.\cite{Giovannetti:prb07} Likewise, LDA describes the interaction between graphene and metal surfaces reasonably well.\cite{Giovannetti:prl08,Khomyakov:prb09,Khomyakov:prb10} We expect a comparably good  description of the interaction between \BN{} and metal surfaces and use the LDA functional throughout.


\begin{figure}
\includegraphics[scale=0.6]{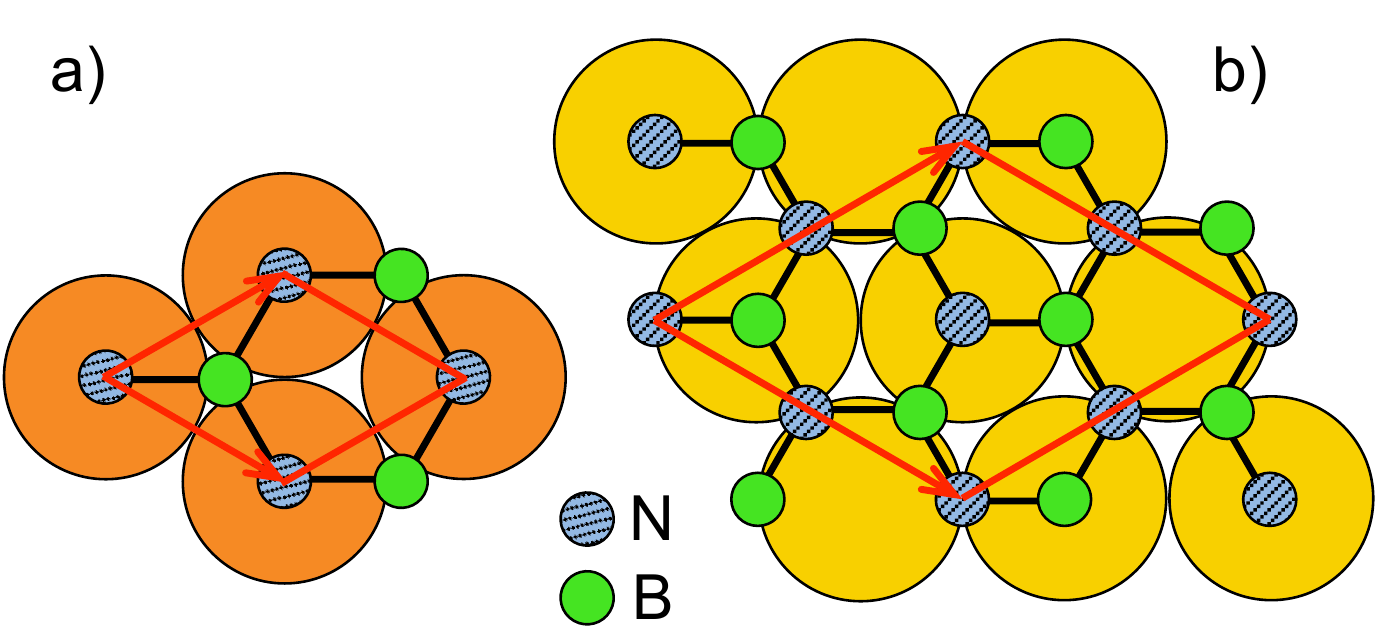}
\caption{(color online) Top views of the energetically favorable configurations of \BN{} on a) Cu, Ni(111), Co(0001), and b) Pt, Pd, Au, Ag, Al (111) in 1$\times$1 and $\sqrt{3} \times \sqrt{3} $  surface unit cells, respectively.} \label{fig-1x1-2x2}
\label{top-cell}
\end{figure}

\begin{figure}
\includegraphics[scale=0.4]{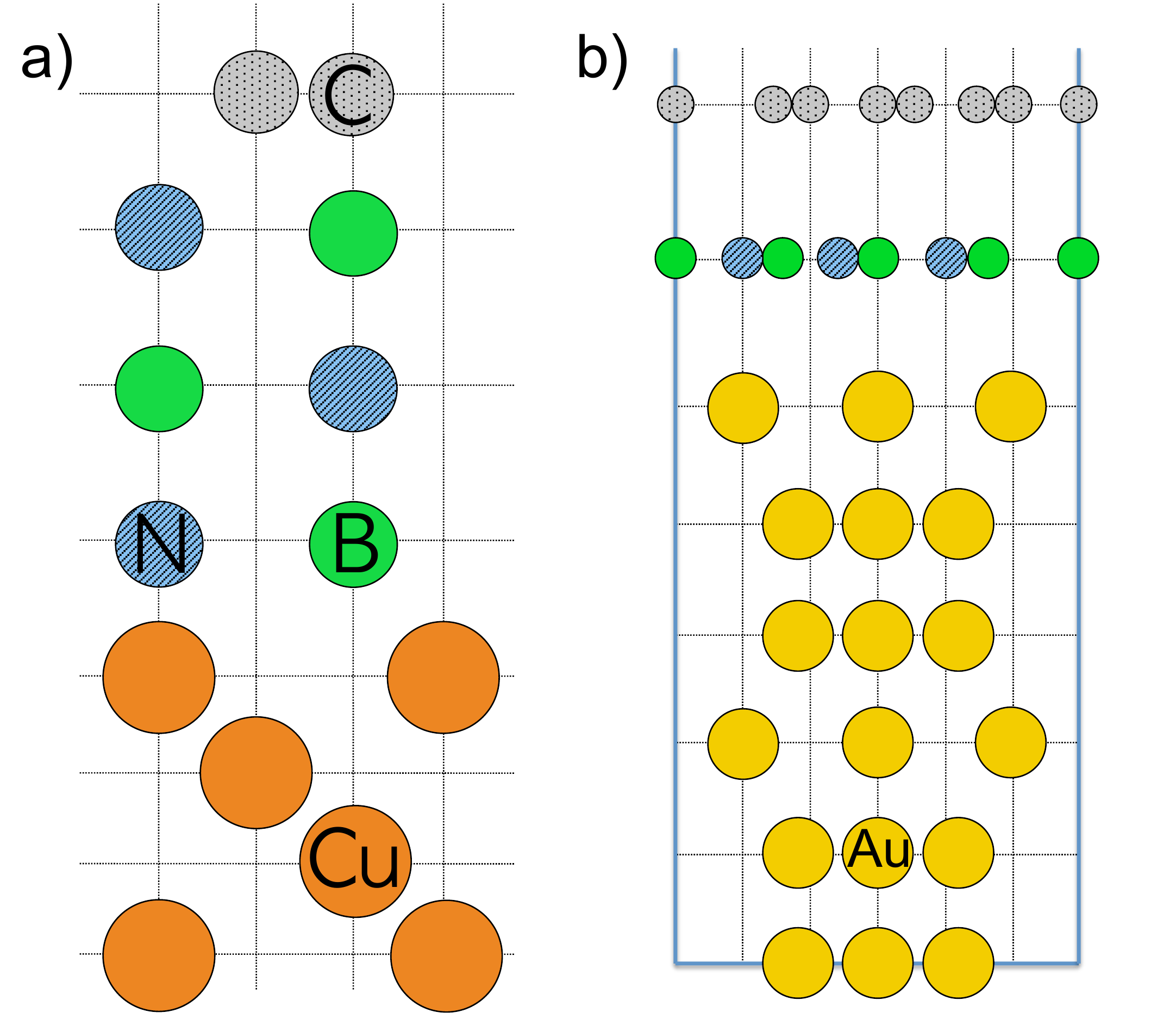}
\caption{(color online) Side views of the energetically favorable configurations of metal$|$\BN{}$|$graphene structures, a) Cu, Ni, Co and b) Au, Pt, Ag, Pd, Al.}  
\label{MBNC-cell}
\end{figure}


For metal$|$\BN{} interface calculations, we fixed the lattice constant of \BN{} to its optimized LDA value of $a_{\rm hex}=2.49$ \AA{}. The in-plane lattice parameter of the metal surfaces was then adjusted so that for Cu, Ni and Co we could use a common 1$\times$1 surface unit cell, Fig.~\ref{top-cell}a. For Au, Pt, Al, Ag, and Pd, a $\sqrt{3} \times \sqrt{3} $ surface unit cell matches very well to a 2$\times$2 surface unit cell of \BN{}, Fig.~\ref{top-cell}b. Adopting this to model the interface involves changing the in-plane lattice parameter of the metals by less than 2\% compared to the optimized values. Such a change does not greatly perturb the electronic structure of the metals, especially after allowing the top  two metal layers to relax in a direction perpendicular to the surface. The small change in the metal lattice parameter is found to have only a small effect on the properties we are interested in. For instance, the work function of the Cu(111) surface is increased by a mere 0.05 eV and the density of states at the Fermi level is unaltered.

In the most stable configuration of a \BN{} sheet on Cu, Ni and Co, nitrogen atoms are adsorbed on top of metal surface atoms and boron atoms are at hollow sites as shown schematically in Fig.~\ref{top-cell}a. The most stable structure for \BN{} on Au, Pt, Al, Ag, and Pd is shown in Fig.~\ref{top-cell}b. In an interface unit cell containing 3 surface metal atoms, 4 boron and 4 nitrogen atoms, one boron and one nitrogen atom is absorbed on top of a metal surface atom; all other boron and nitrogen atoms are on bridge sites. These structures are used to calculate the metal$|$\BN{} interface dipoles.

For calculations on metal$|$\BN{}$|$graphene stacks we use the structures shown schematically in Fig.~\ref{MBNC-cell}. Here we choose the lattice constant of graphene equal to its optimized LDA value $a = 2.445$ \AA, and adapt the in-plane lattice constants of \BN{} and the metal accordingly. This means that the \BN{} is compressed in plane by 1.8\% and the metal by up to 3.8\%. The same procedure was used in our earlier studies of graphene on metal surfaces\cite{Khomyakov:prb09} and on bulk \BN{},\cite{Giovannetti:prb07} the basic rationale being that a change of the in-plane lattice parameter of the metal or of \BN{} will have a smaller effect on their electronic structures (at the Fermi energy) than a corresponding change in graphene would have on its properties. For instance, compressing the Cu(111) lattice by 4\% increases its work function by only 0.08 eV. As a second example, the potential step at the Cu(111)$|$\BN{} interface is $\Delta_\mathrm{Cu|BN}=$1.12 eV for $a_{\rm hex}=2.445$\AA{}, the lattice constant of graphene, and 1.18 eV for $a_{\rm hex}=2.49$\AA{}, the \BN{} lattice constant.

\BN{} layers are stacked in AA$'$ fashion, with boron on top of nitrogen and vice versa, see Fig.~\ref{MBNC-cell}a. The graphene layer is placed on top of \BN{} as in Ref.~\onlinecite{Giovannetti:prb07}. The optimized LDA interlayer separation in the \BN{} slab is 3.24 \AA, which is similar to what is found in bulk \BN{}.  For the Au(111)$|$monolayer \BN{}$|$graphene structure (Fig.~\ref{MBNC-cell}b) the calculated equilibrium separation between the top \BN{} layer and graphene is 3.21 \AA, which is similar to the calculated separation of 3.22 \AA\ between graphene and bulk \BN{}, \cite{Giovannetti:prb07} as expected.

\begin{figure}
\includegraphics[scale=0.46]{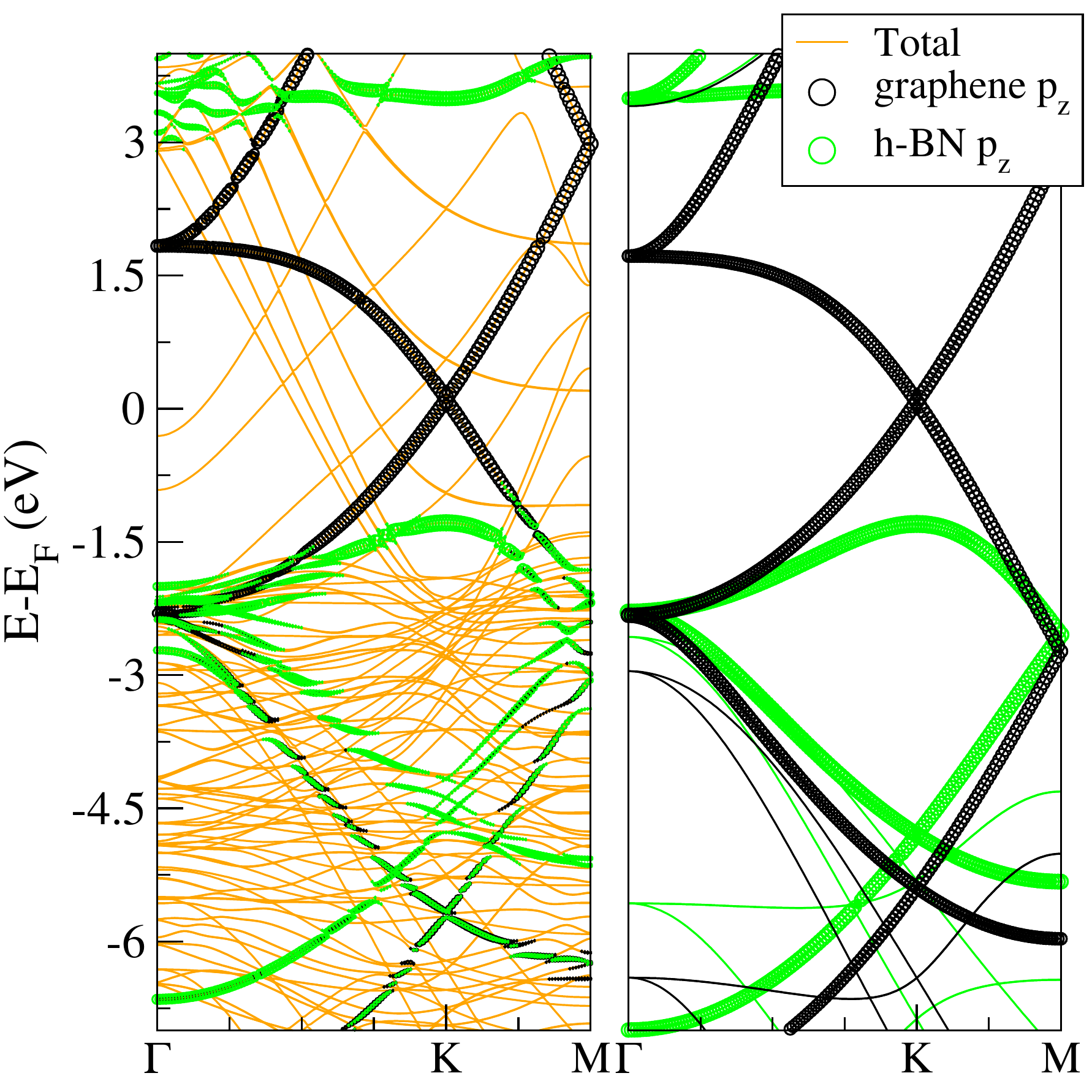}
\caption{(color online) Left panel: the band structure of a Au(111)$|$\BN{} monolayer$|$graphene stack. The  eigenstates of graphene and \BN{} with strong $\pi$ character are highlighted in black and green (gray), respectively where the symbol sizes represent the amount of $p_z$ character. The plotting order is total, graphene $p_z$, \BN{} $p_z$. Right panel: the superimposed band structures of free-standing graphene and a \BN{} monolayer. The graphene (\BN{}) bands have been shifted up (down) by 0.08 (1.28) eV.}\label{AuBNC-bands}
\end{figure}

In the analytical model outlined in Sec. \ref{sec:model} we have assumed that, apart from a shift in the Fermi level, the electronic structure of graphene in a metal$|$\BN{}$|$graphene stack is unchanged compared to a free-standing sheet of graphene. We can check this statement explicitly by analyzing the electronic structure obtained from a DFT calculation on a M$|$\BN{}$|$Gr stack. As an example, the left panel of Fig.~\ref{AuBNC-bands} shows the band structure of a Au(111) substrate with a monolayer of \BN{} and graphene on top. The graphene and \BN{} $p_z$ character of the eigenstates is indicated to help identify the bands. For comparison, the electronic structures of free-standing graphene and of a \BN{} monolayer are shown superimposed in the right panel of Fig.~\ref{AuBNC-bands}. To make the Dirac points of graphene coincide in energy, the band structure of free standing graphene in the right panel has been shifted downwards by 0.08 eV corresponding to the doping level of graphene in the Au$|$\BN{} monolayer$|$graphene structure. The top of the \BN{} valence band, at K, is positioned at $-1.28$ eV, corresponding to the Schottky barrier for holes found for the Au$|$\BN{} structure. Apart from these energy shifts, the band structures of both graphene and \BN{} in the Au$|$\BN{}$|$graphene stack are hardly changed with respect to their free-standing counterparts. 

This reflects the weakness of the interaction between Au and \BN{}, and between \BN{} and graphene. Some metal substrates have a stronger interaction with \BN{} which does affect the \BN{} derived bands. However, as far as the doping levels of graphene are concerned, the effects of that interaction can be fully incorporated in the potential step $\Delta_\mathrm{M|BN}$ associated with the metal$|$\BN{} interface that will be discussed in the next section. If the \BN{} and the graphene lattices are assumed to be commensurate, the sublattice-symmetry-breaking potential of \BN{} induces a small band gap of $\sim 54$ meV between the tips of the Dirac cones;\cite{Giovannetti:prb07} apart from the formation of this small gap, the DoS remains linear in energy. If the lattices are incommensurate, this gap disappears.\cite{Xue:natm11,Dean:natn10,Decker:nanol11,Yankowitz:natp12} In the following this point will not be important.

\section{Results}\label{sec:results}

\begin{figure} 
\includegraphics[scale=0.36]{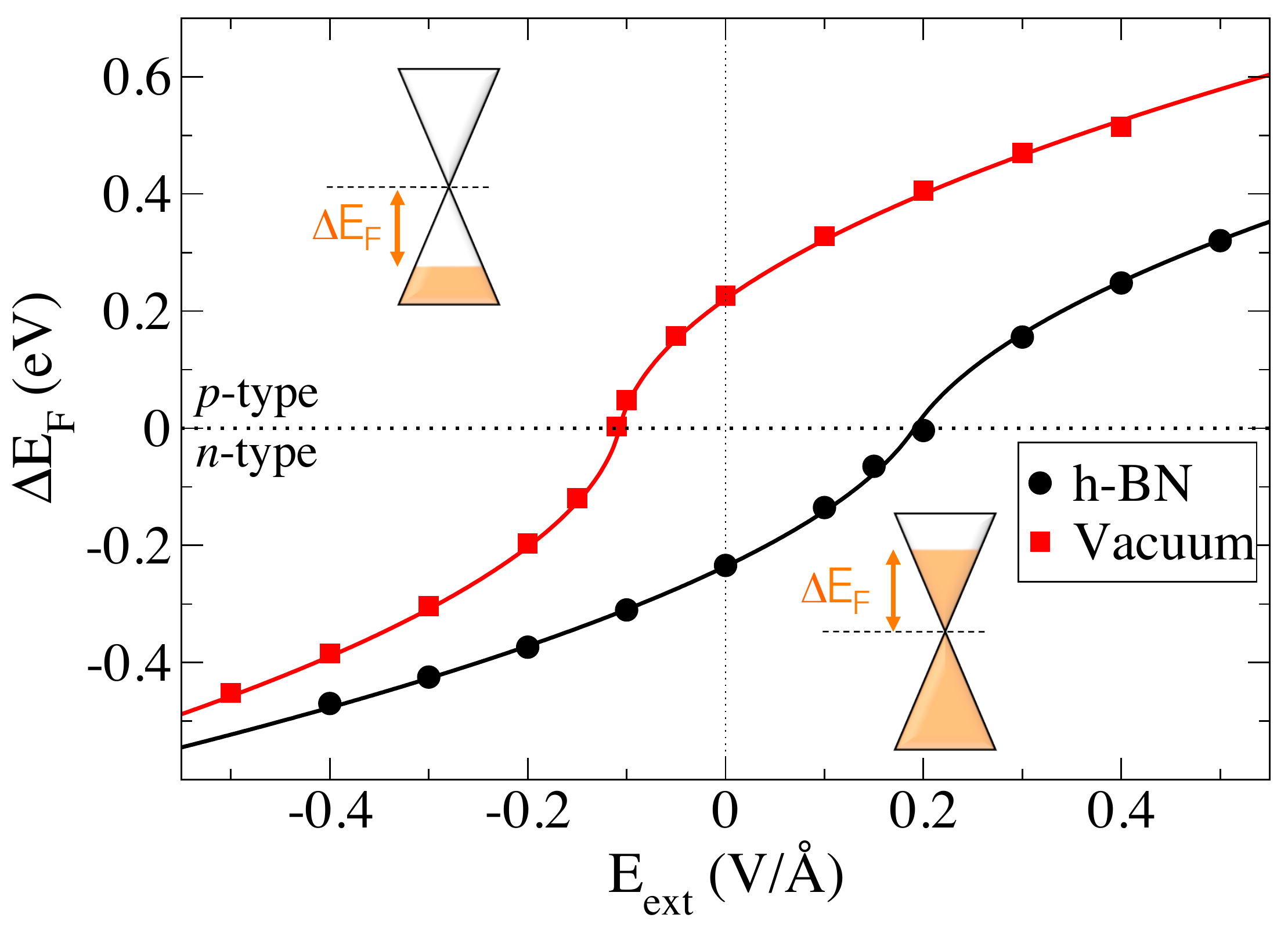}
\caption{(color online) The Fermi level shift versus external electric field strength for a Cu(111)$|$\BN{}$|$graphene structure with two layers of \BN{} (black) and for the same structure with the \BN{} layers replaced by vacuum (red). The lines represent the model described by Eq.(\ref{fermishiftE}) with $\kappa = 2.80$ and 1.0, respectively. The symbols are the values calculated with DFT. } 
\label{CuvacC}
\end{figure}

Conceptually, the simplest model we can consider for a study of electrostatic doping of graphene based upon first-principles calculations is a metal$|$vacuum$|$graphene stack. Fig.~\ref{CuvacC} shows the shift of the Fermi level in graphene calculated using Eq.~(\ref{eq1}), as a function of the external electric field for the Cu(111)$|$vacuum$|$graphene heterostructure, where the vacuum ``thickness'' $d=7.18$ \AA. The points in this figure are the calculated DFT values,  the solid line represents the model described by Eqs.~(\ref{fermishiftE}) and (\ref{E0}) with  $\kappa=1$ and $\Delta_{\rm M|D} = \Delta_{\rm D|Gr} =0$ for the vacuum dielectric. The match between the model (solid curve) and the DFT values (squares) is excellent. 

At zero external bias, the work function difference between the Cu(111) surface and graphene sheet transfers electrons from one to the other until equilibrium is established. The transferred charge sets up an intrinsic electric field $E_0$ that is determined by the work function difference, see Eq.~(\ref{E0}). The work functions of the clean Cu(111) surface and free-standing graphene calculated in the LDA are 5.25 eV and 4.48 eV, respectively, which indicates that at zero external bias electrons are transferred from graphene to Cu so that graphene is doped $p$-type. A compensating \emph{negative} external field can be applied to undo the graphene doping  so $\Delta E_\mathrm{F}=0$. The values given by the model of the compensating field and the Fermi level shift at zero bias are in excellent agreement with the DFT results on the  Cu(111)$|$vacuum$|$graphene system, see Fig.~\ref{CuvacC}.

Insertion of a \BN{} dielectric spacer between the Cu and graphene yields a system that can be realized in practice. The Fermi level shift in graphene for a Cu(111)$|$\BN{}$|$graphene structure with two layers of \BN{} (corresponding, by construction, precisely to the vacuum thickness used in the previous calculation) is also shown in Fig.~\ref{CuvacC} as a function of the external electric field. Comparing the results obtained with a \BN{} spacer and a vacuum spacer, two qualitative differences are immediately obvious. The curve calculated with \BN{} results is shifted to the right, and it is ``flatter''. 

The latter effect results largely from the dielectric constant of the spacer;  Eq.~(\ref{fermishiftE}) shows that $\Delta E_\mathrm{F}$ decreases with increasing dielectric constant. We determined the dielectric constant of \BN{} from calculations on  very thin \BN{} films as the ratio between an applied external electric field and the internal electric field inside the slab \cite{Pham:apl10} determined from the shift of the 1s Nitrogen core levels. Because these levels are localized on each layer, their energies shift according to the local electrostatic potential. The shift is linear in the position of the layer, and the slope yields the internal electric field. From calculations on \BN{} slabs consisting of 2-10 layers in electric fields in the range 0.05-0.45 V/\AA, we find $\kappa=2.80\pm0.02$. In our modelling in the remainder of this paper, we use $\kappa=2.80$. Fig.~\ref{CuvacC} shows that with this value of $\kappa$ the shape of the curve gives an excellent description of the DFT results for the Cu(111)$|$\BN{}$|$graphene heterostructure.  

The shift between the two curves in Fig.~\ref{CuvacC} results from potential steps $\Delta_{\rm M|D}$ and $\Delta_{\rm D|Gr} $ that are formed at the metal$|$\BN{} and \BN{}$|$graphene interfaces, respectively. These potential steps are sufficiently large not only to compensate for the difference in work functions of the Cu(111) surface and free-standing graphene, but even to reverse the intrinsic electric field (Eq.~(\ref{E0})). It means that in the Cu(111)$|$\BN{}$|$graphene stack at zero bias, graphene is $n$-type doped, and one needs to apply a positive external field to undo the graphene doping and enforce $\Delta E_\mathrm{F}=0$. Because the interface potential steps play such an important role, we study them in more detail in the next subsection.

\subsection{Potential steps at metal$|$\BN{} and \BN{}$|$graphene interfaces} 

In general, potential steps are formed at interfaces between different materials. They result from interface dipole layers, caused by chemical interactions and/or charge transfer between two materials comprising an interface. Substantial interface dipoles can be formed even if the interaction between the two materials is relatively weak as happens when noble gas atoms are adsorbed on, or closed-shell molecules are physisorbed on metal surfaces.\cite{Bagus:prl02,Rusu:jpcc09,Rusu:prb10,Bokdam:nanol11} Even in the absence of a direct interaction, the difference between the work functions of two metals leads to charge transfer and the formation of contact potentials.

The potential step $\Delta_\mathrm{M|D}$ at the metal$|$dielectric interface can be obtained by comparing the work function $W_{\rm M}$ of the clean metal surface without the adsorbate to that of the metal with the dielectric adsorbed on it, $W_{\rm M|D}$. Similarly, the potential step $\Delta_\mathrm{D|Gr}$ is given by the difference between the work functions on the dielectric side  and the graphene side of a $\mathrm{D|Gr}$ slab
\begin{equation}
\Delta_{\rm M|D}=W_{\rm M}-W_{\rm M|D};\,\;
\Delta_{\rm D|Gr}=W_{\rm D|Gr}-W_{\rm Gr|D}.
\label{dc}
\end{equation}
It turns out to be sufficient to study graphene adsorbed on two layers of \BN{} to find a converged value for $\Delta_{\rm BN|Gr}=0.14$ eV. Similarly, a single \BN{} layer adsorbed on a metal substrate is sufficient to determine $\Delta_{\rm M|BN}$, demonstrating the true interface nature of these potential steps. The values of $\Delta_{\rm M|BN}$ calculated for the metals considered in this paper, given in Table~\ref{tbMBN}, are seen to be quite substantial, ranging from 0.4 eV for Al(111)$|$\BN{} to 1.8 eV for Co(0001)$|$\BN{}. In all cases $\Delta_{\rm M|BN}$ is positive, meaning that adsorption of \BN{} on a metal substrate lowers the work function. In other words, the interaction between \BN{} and the metal substrate leads to a net displacement of the electronic charge distribution {\em into} the metal.

The metal$|$\BN{} equilibrium distance $d_{\rm eq} $ indicates that there are two different adsorption regimes. Adsorption of \BN{} on the Co(0001), Ni(111) and Pd(111) surfaces leads to equilibrium distances $d_{\rm eq}<2.5$ \AA,  values which are typically associated with chemisorption, whereas the values of $d_{\rm eq}>3.0$ \AA\ for adsorption on the Cu, Pt, Ag, Au, and Al(111) surfaces are typical of physisorption.

The calculated values of $\Delta_{\rm M|BN}$ increase with decreasing separation of the \BN{} plane from the metal substrate, as shown in Table~\ref{tbMBN}.  A smaller separation leads to an increased metal$|$\BN{} interaction, and hence to a larger deformation of the charge density and a larger potential step at the interface. The calculated $\Delta_{\rm M|BN}$ generally agree well with the experimental values. However, for ${\rm Cu|BN}$, different experiments report quite different results.\cite{Preobrajenski:ss05,Joshi:nanol12} The values we calculate are in better agreement with values recently reported in Ref.~\onlinecite{Joshi:nanol12}.  In view of the results shown in Table~\ref{tbMBN}, the experimental value $\Delta_{\rm Cu|BN} = 0.24$ eV reported in Ref. \onlinecite{Preobrajenski:ss05} is strikingly low.

\begin{table} 
\caption{Calculated and experimental potential steps $\Delta_{\rm M|BN}$ at the metal$|$\BN{} interface. $d_{eq}$ is the equilibrium separation of the \BN{} layer from the metal surface. The \BN{} lattice constant $a_\mathrm{hex}=2.49$ \AA\ is used for the surface unit cells, except in the last two rows, where the graphene lattice constant $a_\mathrm{hex}=2.445$ \AA\ is used.}
\begin{ruledtabular}

\begin{tabular}{lcccc}
 M  & $\Delta_{\rm M|BN}$(eV)  &  $\Delta_{\rm M|BN}^{\rm exp} $(eV)   & $d_{\rm eq} $ (\AA)     &  $W_{\rm M} $ (eV) \\
 \hline
 Co   & 1.80 &                        & 1.92 & 5.52 \\
 Ni   & 1.73 & 1.5-1.8$^{a,c}$        & 1.96 & 5.52 \\
 Pd   & 1.25 & 1.3$^b$                & 2.47 & 5.53 \\
 Cu   & 1.18 & 0.24$^c$,0.8-1.1$^{d}$ & 2.97 & 5.17 \\
 Pt   & 1.04 & 0.9$^b$                & 3.04 & 5.98 \\
 Ag   & 0.83 &                        & 3.20 & 4.83 \\
 Au   & 0.79 &                        & 3.24 & 5.55 \\
 Al   & 0.41 &                        & 3.55 & 4.25 \\
 Gr   & 0.11 &                        & 3.21 & 4.63 \\
  \hline
 Gr$'$& 0.14 &                        & 3.22 & 4.48 \\
 Cu$'$& 1.12 & 0.24$^c$,0.8-1.1$^{d}$ & 3.11 & 5.25  \\
\end{tabular} 
\end{ruledtabular}
 \newline
$^a$Refs. \onlinecite{Nagashima:ss96,Grad:prb03,Leuenberger:prb11}, $^b$Ref. \onlinecite{Nagashima:ss96}, $^c$Ref. \onlinecite{Preobrajenski:ss05}, $^d$Ref. \onlinecite{Joshi:nanol12}.
\label{tbMBN}
\end{table}

An interface potential step can have different origins. Chemical interactions between two materials can result in an ordered interface layer of bond dipoles.\cite{Rusu:prb06,Otalvaro:jpcc12}  Even when the interaction is weak, as in the case of physisorption, it is argued that the Pauli exchange interaction between the adsorbate and the substrate can distort the charge density distribution significantly and drive a net displacement of charge towards the ``softer'' material, in this case the metal substrate, which lowers the work function. This is commonly called the ``pushback'' or ``pillow'' effect.\cite{Bagus:prl02} 

In addition to a potential step induced by a direct interaction at the interface, one can have a contribution originating from an electron transfer to equilibrate the Fermi level between the two materials on either side of the interface. Such a transfer obviously occurs only if the Fermi level is in a range where the density of states of both materials is non-zero. At interfaces between a metal and an organic molecular material, where charge is exchanged between strong donor or acceptor molecules and the metal surface, this leads to pinning of the Fermi level by the molecules.\cite{Bokdam:apl11}

One does not expect this to be the case for \BN{} adsorbed on a metal surface. \BN{} is a large band gap insulator, and in most cases the Fermi level of the metal is well within the \BN{}  band gap. The potential steps $\Delta_{\rm M|BN}$ and $\Delta_{\rm BN|Gr}$ can therefore be ascribed to the direct interaction between the materials at the interface. 

We can visualize the electronic displacement and dipole formation at the interface by looking at the electron density of the entire system minus the electron densities of the two separate constituent materials. As only the component perpendicular to the interface is relevant in the present context, it is convenient to work with plane-averaged electron densities $n(z)=\int n(x,y,z)dxdy/A$ in terms of which the electron charge density $\rho = -en$. The electron displacement in a metal$|$\BN{} structure is defined as
\begin{eqnarray}\label{n1}
\nonumber 
\lefteqn{\Delta n_{\rm M|BN}(z,E_{\rm ext}) =}\\ 
  & & n_{\rm M|BN}(z,E_{\rm ext})-n_{\rm M}(z,E_{\rm ext})-n_{\rm BN}(z,E_{\rm ext}),
\end{eqnarray}
where $n_{\rm M|BN}$ is the electron density of the composite system, $n_{\rm M}$ and $n_{\rm BN}$ are the electron densities of the clean metal substrate and the isolated \BN{} layer, respectively, and we introduce an explicit $E_{\rm ext}$ dependence because we wish to study the dependence of the interface electronic displacement on the external electric field. Any electronic displacement $\Delta n(z)$ can be related to a potential (energy) step $\Delta$ by simple electrostatics, i.e., $\Delta = -e^2\int z \Delta n(z) dz / \epsilon_0$. 

\begin{figure}
\includegraphics[scale=0.38]{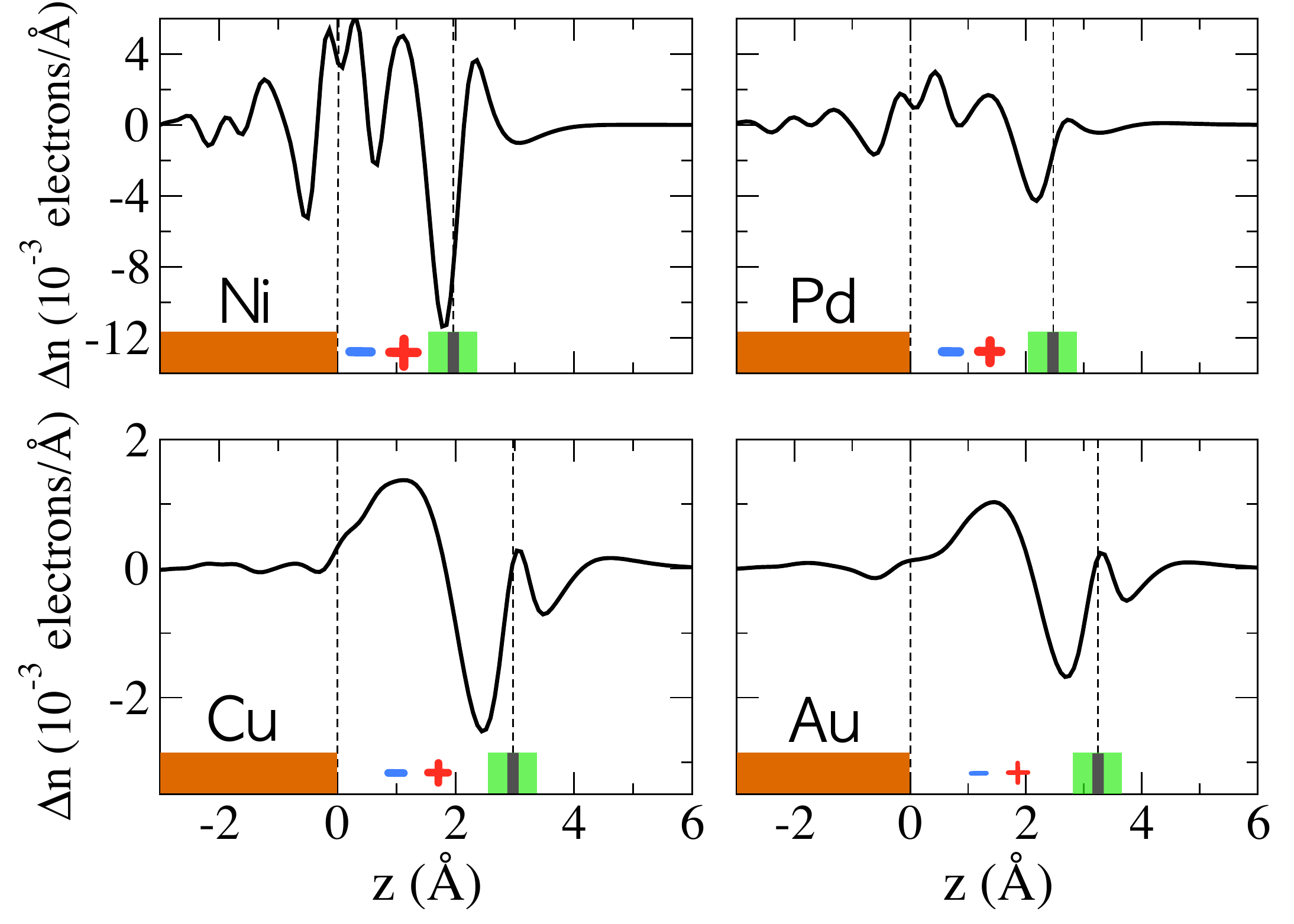}
\caption{(color online) Plane-averaged electron density difference $\Delta n_{\rm M|BN}(z,0)$ for Ni, Pd, Cu and Au (111) surfaces covered with a monolayer of \BN{}. The vertical dashed lines represent the positions of the atomic layers at the interfaces. The position $z$ is measured from the top metal atom.} 
\label{fig:dipMBN}
\end{figure}

The calculated electronic displacement without an external field, $\Delta n_{\rm M|BN}(z,E_{\rm ext}=0)$, is shown in Fig.~\ref{fig:dipMBN} for four different metal$|$\BN{} interfaces. The amplitude of $\Delta n_{\rm M|BN}(z)$ is seen to decrease with increasing separation of the adsorbed \BN{} layer.  The oscillation patterns illustrate the two different kinds of behavior discussed above. For \BN{} adsorbed on the Cu(111) and Au(111) surfaces, the pattern of $\Delta n_{\rm M|BN}(z)$ in the region between the metal surface and the \BN{} layer is that of a simple dipole. It corresponds to an accumulation of electrons close to the metal surface and a depletion close to the \BN{} layer, consistent with the pillow effect discussed above. The simple interface dipole and the large metal$|$\BN{} equilibrium separations (Table~\ref{tbMBN}) found for these systems are in line with physisorption of \BN{} on these metals. In contrast, the oscillation pattern of $\Delta n_{\rm M|BN}(z)$ for \BN{} adsorbed on Ni(111) and Pd(111) surfaces is much more complicated. It is accompanied by a small metal-\BN{} equilibrium distance, which is consistent with chemisorption.

\begin{figure}[b]
\includegraphics[scale=0.36]{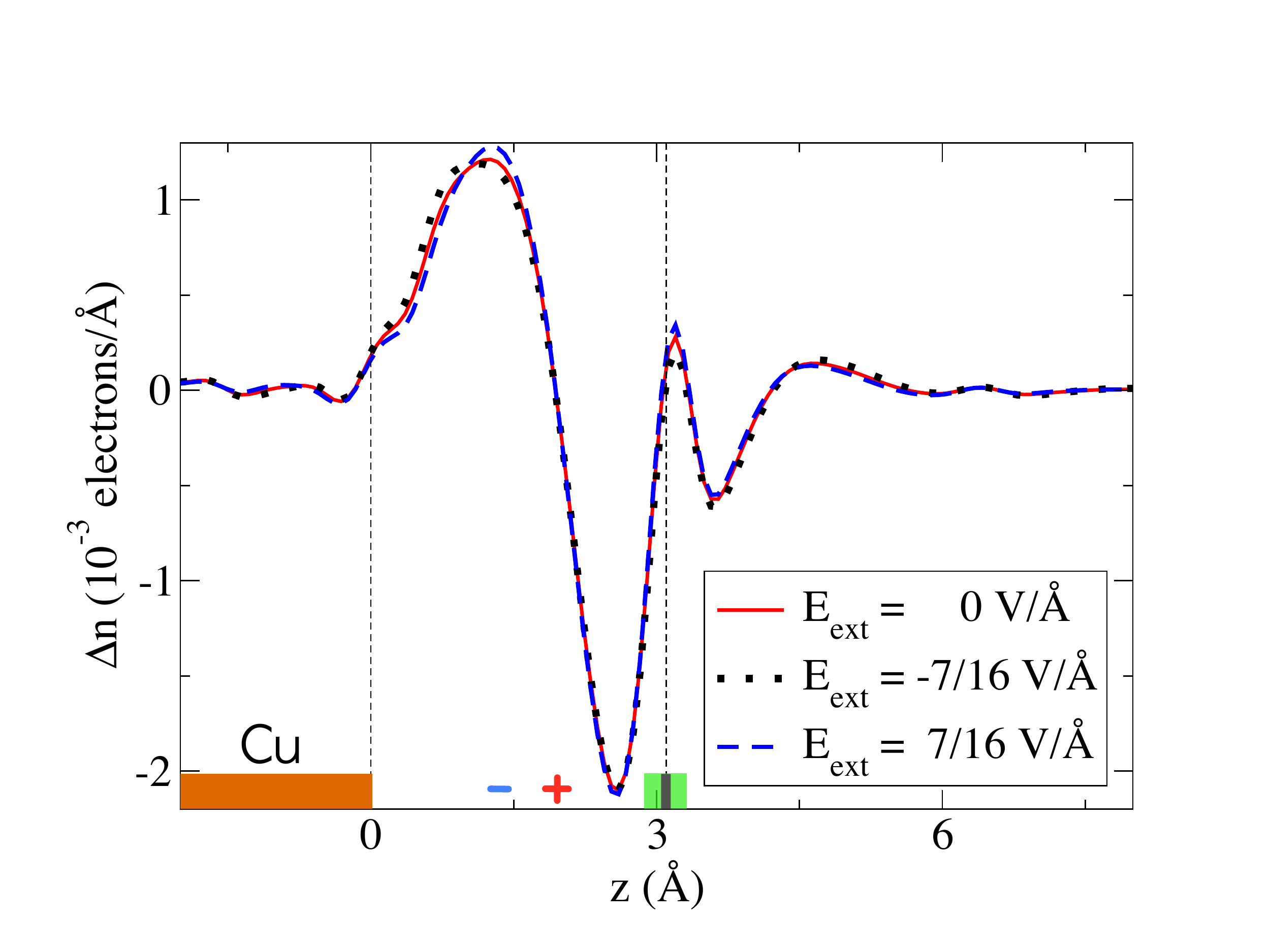}
\caption{(color online) Influence of an external electric field on the Cu(111)$|$BN interface dipole layer. The plane-averaged electron density difference $\Delta n_{\rm Cu|BN}(z,E_{\rm ext})$ is shown for three different electric fields (perpendicular to the interface). The vertical dashed lines represent the positions of the atomic layers at the interface.} 
\label{E-depCuBN}
\end{figure}

\begin{figure}
\includegraphics[scale=0.36]{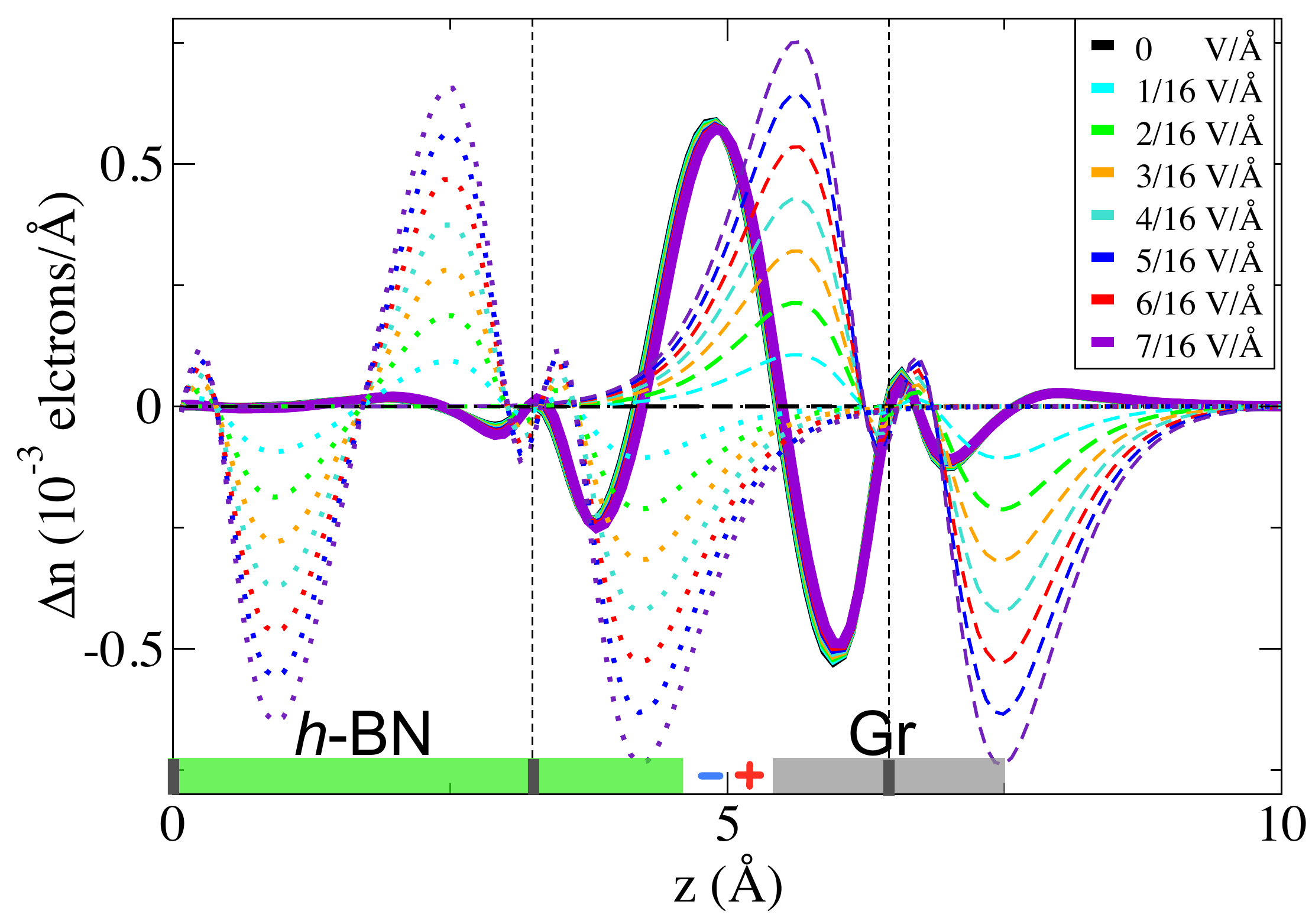}
\caption{(color online) Influence of an external electric field on the BN$|$Gr interface dipole layer. The plane-averaged electron density difference $\Delta n_{\rm BN|Gr}(z,E_{\rm ext})$ is shown for different electric field strengths (solid lines). The dotted and dashed lines show the polarization of isolated \BN{} and Gr sheets, respectively, in these external fields. The vertical dashed lines represent the positions of the atomic layers at the interfaces.} 
\label{E-depBNC}
\end{figure}

In the model described by Eqs.~(\ref{fermishiftE})-(\ref{eq:V0}), it was assumed that the interface potential steps $\Delta_{\rm M|D}$ and $\Delta_{\rm D|Gr}$ are independent of an applied external field. That implies that the corresponding interface dipoles should be independent of the external field. Figure \ref{E-depCuBN} shows  $\Delta n_{\rm Cu|BN}(z,E_{\rm ext})$ for \BN{} adsorbed on the Cu(111) surface calculated with $E_{\rm ext}=-0.44,$ $0$ and $+0.44$ V/\AA. This figure clearly demonstrates that  $\Delta n_{\rm Cu|BN}(z,E_{\rm ext})$ is essentially independent of the electric field strength within this range, and therefore also the interface dipole and the potential step. A similar independence can be observed in the electronic displacement at a \BN{}$|$graphene interface in Fig.~\ref{E-depBNC} (solid lines). Hence the interface dipoles, and the potential steps $\Delta_\mathrm{M|BN}$ and $\Delta_\mathrm{BN|Gr}$ are truly intrinsic properties of these interfaces.

The fact that the electronic displacement at the interfaces is field-independent (for the field strengths used here) does not mean that the electron densities are so, as both \BN{} and graphene are polarized in an electric field. Fig. \ref{E-depBNC} shows the changes in electron density $n(z,E_{\rm ext})-n(z,0)$ upon applying an electric field, both for an isolated graphene layer (dashed lines) and for an isolated \BN{} slab (dotted lines). The changes clearly illustrate the induced dipoles on the individual layers.

\begin{figure}
\includegraphics[scale=0.36]{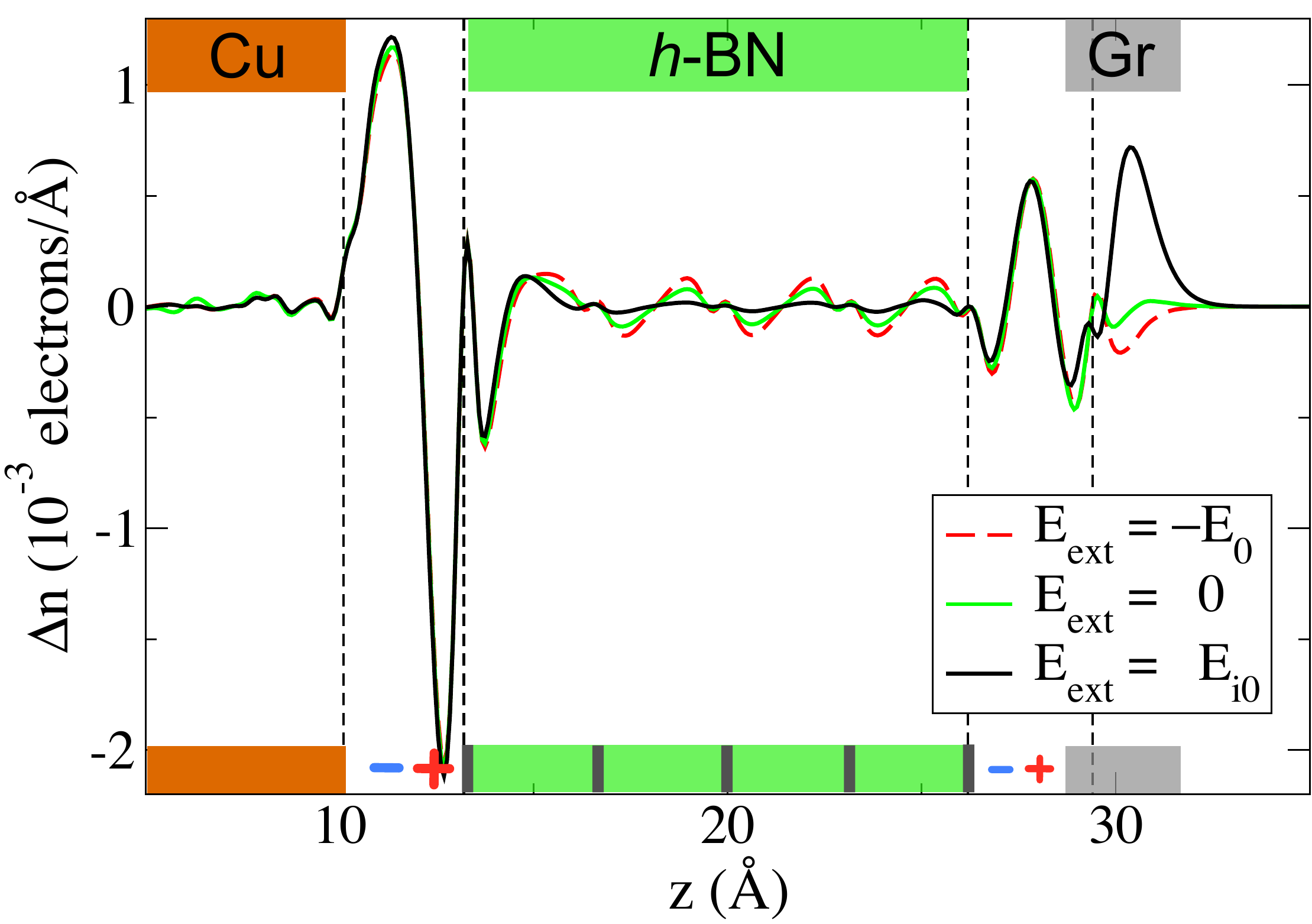}
\caption{(color online) Plane-averaged electron density difference $\Delta n_{\rm tot}(z,E_{\rm ext})$ for a Cu(111)$|$5-layers-of-\BN{}$|$graphene structure for external electric fields $E_{\rm ext}=-E_{\rm 0},0,E_{\rm i0}$ V/\AA. The vertical dashed lines represent the positions of the atomic layers at the interfaces. } \label{fig-p3}
\end{figure}

For a complete metal$|$\BN{}$|$graphene stack, we visualize the total electronic displacement caused by interface formation, as well as by the external electric field,  as
\begin{eqnarray}\label{n3}
 &\Delta& n_\mathrm{tot}(z,E_{\rm ext})= \nonumber \\
&n&_{\rm M|BN|Gr}(z,E_{\rm ext})-n_{\rm M}(z)-n_{\rm BN}(z)-n_{\rm Gr}(z),
\end{eqnarray}
where the reference charge densities $n_{\rm M}$, $n_{\rm BN}$, $n_{\rm Gr}$ are for the isolated systems in zero external field. Results are shown in Fig.~\ref{fig-p3} for a Cu(111)$|$\BN{}$|$graphene structure with five layers of \BN{}. The dipoles at the Cu(111)$|$\BN{} and \BN{}$|$graphene interfaces are essentially the same as the dipoles at  individual isolated interfaces, see Figs.~\ref{E-depCuBN} and \ref{E-depBNC}, implying that the two interfaces in the Cu(111)$|$\BN{}$|$Gr stack are not coupled.
	
$\Delta n_\mathrm{tot}$ in Fig.~\ref{fig-p3} for the zero external field case ($E_{\rm ext}=0$) shows small oscillations in the \BN{} slab, indicating that the \BN{} layers are polarized. The effective work function difference $V_{\rm 0}$ between the Cu substrate and graphene, Eq.~(\ref{eq:V0}), results in electron transfer from Cu to graphene across the \BN{} slab to establish a single Fermi level in equilibrium. This sets up an intrinsic electric field in the \BN{} slab, see Eq.~(\ref{E0}). This charge transfer can be counteracted by applying an external electric field $E_{\rm ext}=-E_{\rm 0}$, see Eq.~(\ref{fermishiftE}). 

This external field results in charge neutral (i.e. undoped) graphene, but the field in the \BN{} slab is still non-zero, see Eq.~(\ref{eq4}) with $\sigma=0$ and $E_{\rm ext}=-E_{\rm 0}$. An external field of $E_{\rm ext}=E_{\rm i0}\approx-0.43$ V/\AA, Eq. (\ref{Ei0}), gives a zero internal field, see Fig.~\ref{fig-p3}. Note that, whereas $\Delta n_\mathrm{tot}$ in the region between the topmost \BN{} layer and the graphene sheet is almost field-independent, it strongly depends on the field on the graphene side ($z\gtrsim 30$ \AA). This just reflects the polarization behavior of the isolated graphene sheet, see Fig.~\ref{E-depBNC}. 

\subsection{Fermi level shifts in metal$|$\BN{}$|$graphene heterostructures}

All the parameters that appear in Eqs.~(\ref{fermishiftE}-\ref{eq:V0}) have now been calculated by studying constituent parts of M$|$\BN{}$|$Gr stacks, Table~\ref{tbMBN}, and the model can be tested by comparing its predictions to the results obtained from explicit DFT calculations on full M$|$\BN{}$|$Gr structures. Fig.~\ref{CuvacC} demonstrates that the electric field dependence of the Fermi level shift is described very well by the model. We would like to emphasize that these curves do not represent data fits, as all parameters in the model are fixed. 

We next study how the thickness of \BN{}, measured in terms of the number of discrete layers $n$, influences the Fermi level shift  $\Delta E_{\rm F}$ in graphene. In Table \ref{tab-data}, $\Delta E_{\rm F}$ calculated with DFT for Cu(111)$|n\times$\BN{}$|$graphene structures with $0<n<6$ layers is compared to values predicted by the model for a vanishing external field, $\Delta E_{\rm F}^\mathrm{m}$. The agreement between the two sets of results is to within 12 meV or better. The $n=0$ case corresponds to graphene directly adsorbed onto the Cu(111) surface and the results agree with those found previously.\cite{Giovannetti:prl08,Khomyakov:prb09} The good agreement found in Table~\ref{tab-data} indicates that the two interfaces are not strongly coupled. Even for a single \BN{} layer, $n=1$, the model works well.

\begin{table}[t!]
\caption{Calculated work function $W$ and Fermi energy shift $\Delta E_{\rm F}$ of Cu(111)$|n \times $\BN{}$|$graphene structures with $n$ layers of \BN{}, and Fermi energy shifts $\Delta E_{\rm F}^\mathrm{m}$ predicted by the model. All values are in eV.} 
\begin{ruledtabular}
\begin{tabular}{lccccccc}
     $n$            & 0        & 1        & 2        & 3        & 4        & 5        & 6        \\ \hline 
     $W$            & 4.386    & 4.223    & 4.299    & 4.325    & 4.336    & 4.346    & 4.354    \\
 $\Delta E_{\rm F}$ & $-0.171$ & $-0.304$ & $-0.234$ & $-0.206$ & $-0.193$ & $-0.180$ & $-0.171$ \\
 $\Delta E_{\rm F}^{\rm m}$ & $-0.17^{\rm a}$ & $-0.297$ &$-0.246$ & $-0.216$ & $-0.196$ & $-0.181$ & $-0.169$ \\ 

\end{tabular}
\end{ruledtabular}
$^{\rm a}$Ref. \onlinecite{Khomyakov:prb09}
\label{tab-data}
\end{table}

\begin{figure}[b!] 
\includegraphics[scale=0.36]{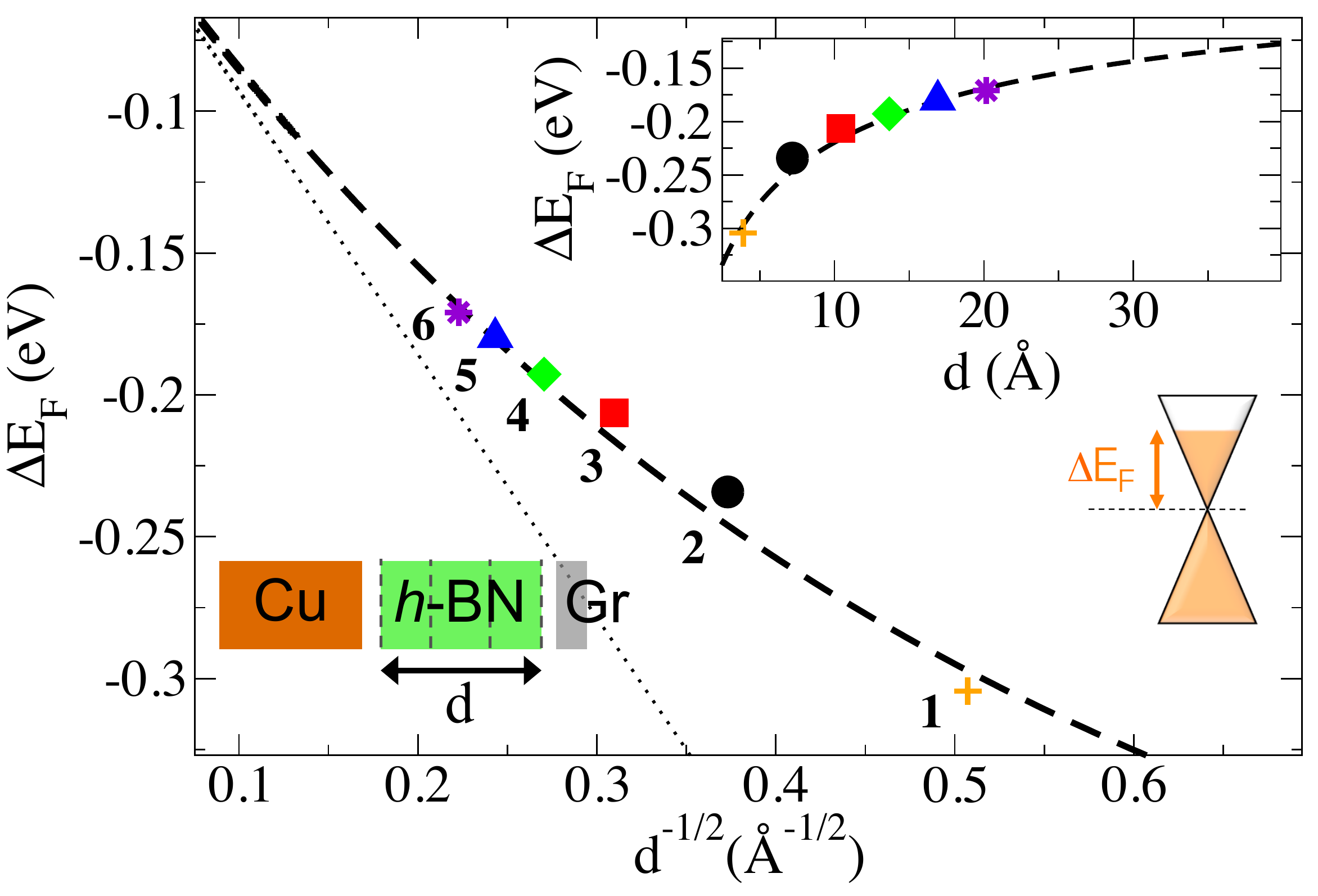}
\caption{(color online) The Fermi level shift versus the thickness $d$ of the \BN{} film, plotted as a function of $1/\sqrt{d}$ and (inset) as a function of $d$. The symbols are the results of DFT calculations on Cu(111)$|$\BN{}$|$graphene structures with 1 - 6 layers of \BN{}. The dashed line represents the model calculated with Eq. (\ref{fermishiftE}), the dotted line represents the asymptotic $1/\sqrt{d}$ dependence.} \label{A}
\end{figure}

In the absence of an applied potential, graphene in a Cu(111)$|$\BN{}$|$graphene stack is doped $n$-type in equilibrium. The degree of doping decreases as the thickness of the dielectric layer is increased. This can be simply understood from Eq.~(\ref{fermishiftE}) where for large values of $d$, the Fermi level shift scales asymptotically with $d$ as $\Delta E_{\rm F} \propto 1/\sqrt{d}$. Fig.~\ref{A} illustrates this scaling, and also demonstrates the deviation of this scaling for small $d$. The good agreement between the results of the first-principles calculations and the model seems to improve with increasing \BN{} thickness.

\begin{figure} [t!]
\includegraphics[scale=0.36]{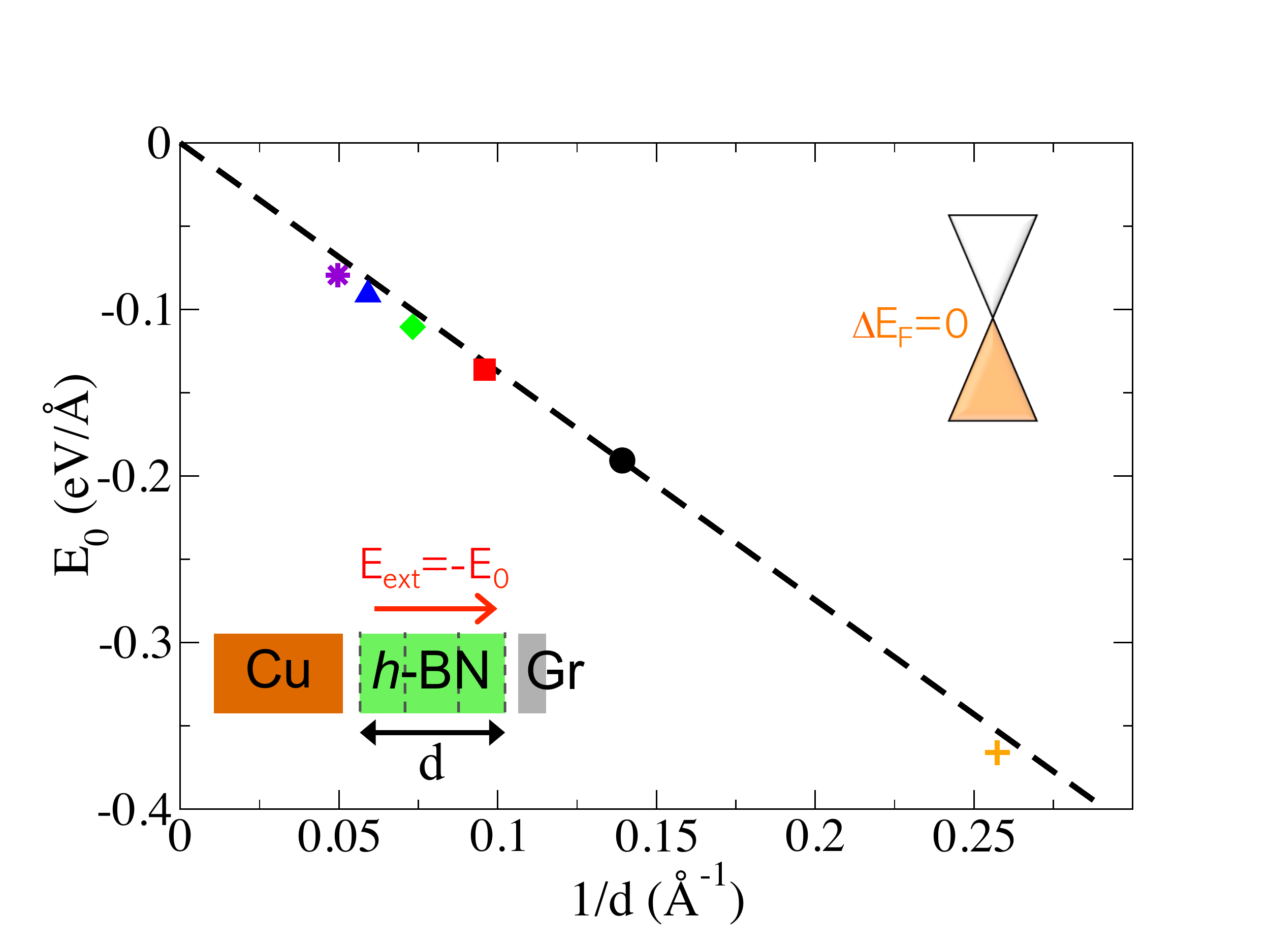}
\caption{(color online) The value of the intrinsic electric field $E_0$ (Eq.~(\ref{E0})) as a function of $1/d$ where $d$ is the thickness of the \BN{} film. An external electric field $E_\mathrm{ext}^0=-E_{\rm 0}$ is required to restore graphene to charge neutrality. The symbols are the results of DFT calculations for Cu(111)$|$\BN{}$|$graphene structures with 1 - 6 layers of \BN{}.} \label{AA}
\end{figure}

As discussed in the previous section, the Fermi level can be positioned at the charge neutrality point of graphene by application of an external electric field $E_\mathrm{ext} = -E_0$. According to  Eq.~(\ref{E0}), $E_{\rm0}$ should scale with the \BN{} layer thickness $d$ as $E_{\rm0} \propto 1/d$. This scaling is demonstrated in Fig.~\ref{AA}. The symbols represent the $E_\mathrm{ext}$ values needed in the DFT calculations to make graphene charge neutral in stacks with variable \BN{} thickness. From Eq.~(\ref{Vg}) we then predict that the gate voltage required to shift the Fermi level to the graphene conical points should be independent of how thick \BN{} is. 

\begin{figure} [b!]
\includegraphics[scale=0.38]{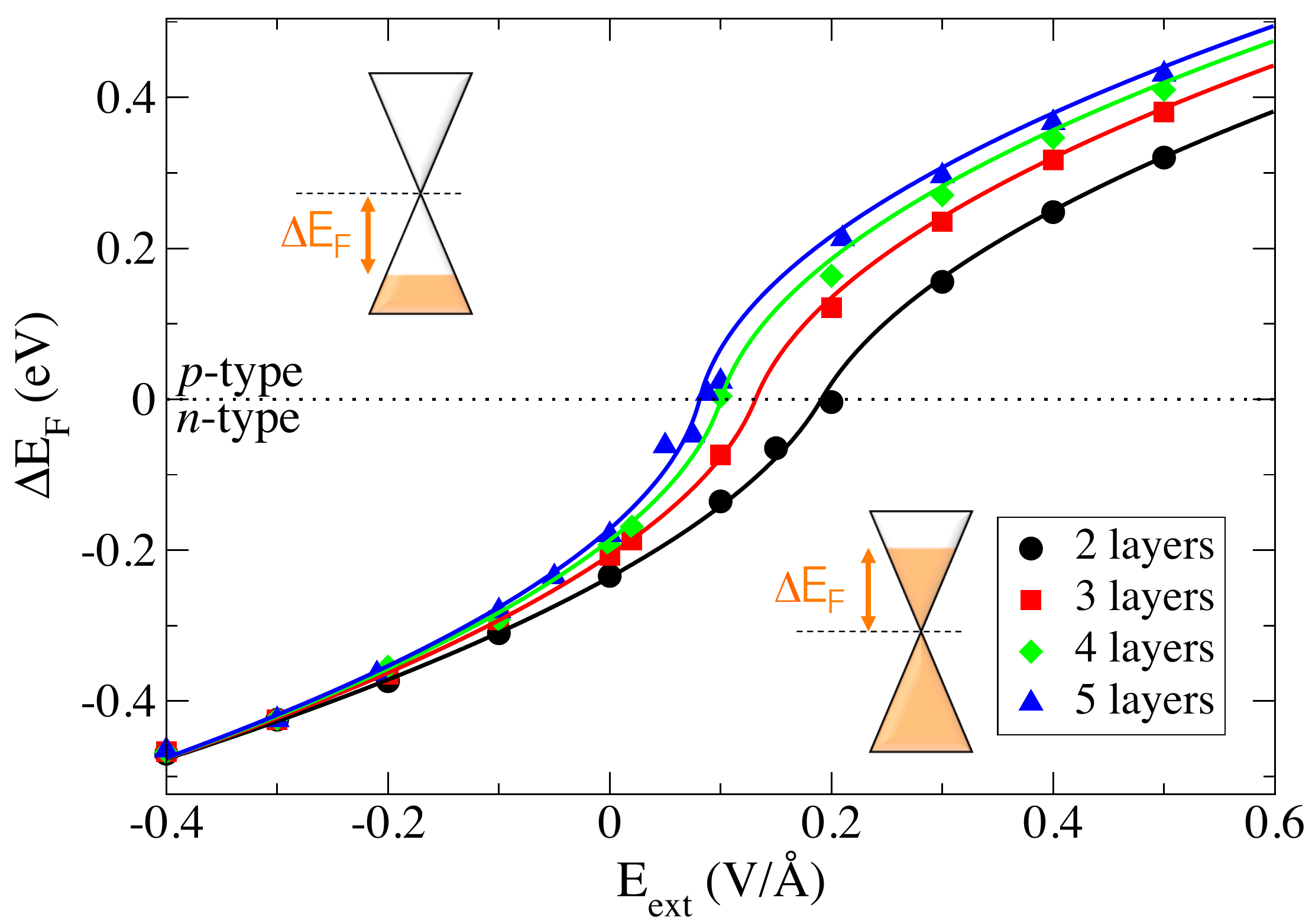}
\caption{(color online) The Fermi level shift versus external electric field strength for a Cu(111)$|$\BN{}$|$graphene structure with 2 - 5 layers of \BN{}. The lines represent the model of Eq.(\ref{fermishiftE}) for different thicknesses of dielectric. The symbols are values of the doping level calculated with DFT. } \label{B}
\end{figure}

The dependence of the doping level, both as a function of the external field $E_{\rm ext}$, and as a function of the \BN{} layer thickness, is shown in Fig.~\ref{B}. The agreement between the model and the DFT results is very good. The characteristic shape of the curves is determined by Eq.~(\ref{fermishiftE}). The curves for different \BN{} layer thicknesses cross at $E_{\rm ext}\approx-0.4$ V/\AA. At this external field strength the doping level is independent of the thickness of the \BN{} dielectric spacer. This point corresponds to the field $E_{\rm i0}=-0.43$ V/\AA\ discussed above, where the internal field in the \BN{} spacer is zero.

\begin{figure}[t] 
\includegraphics[scale=0.36]{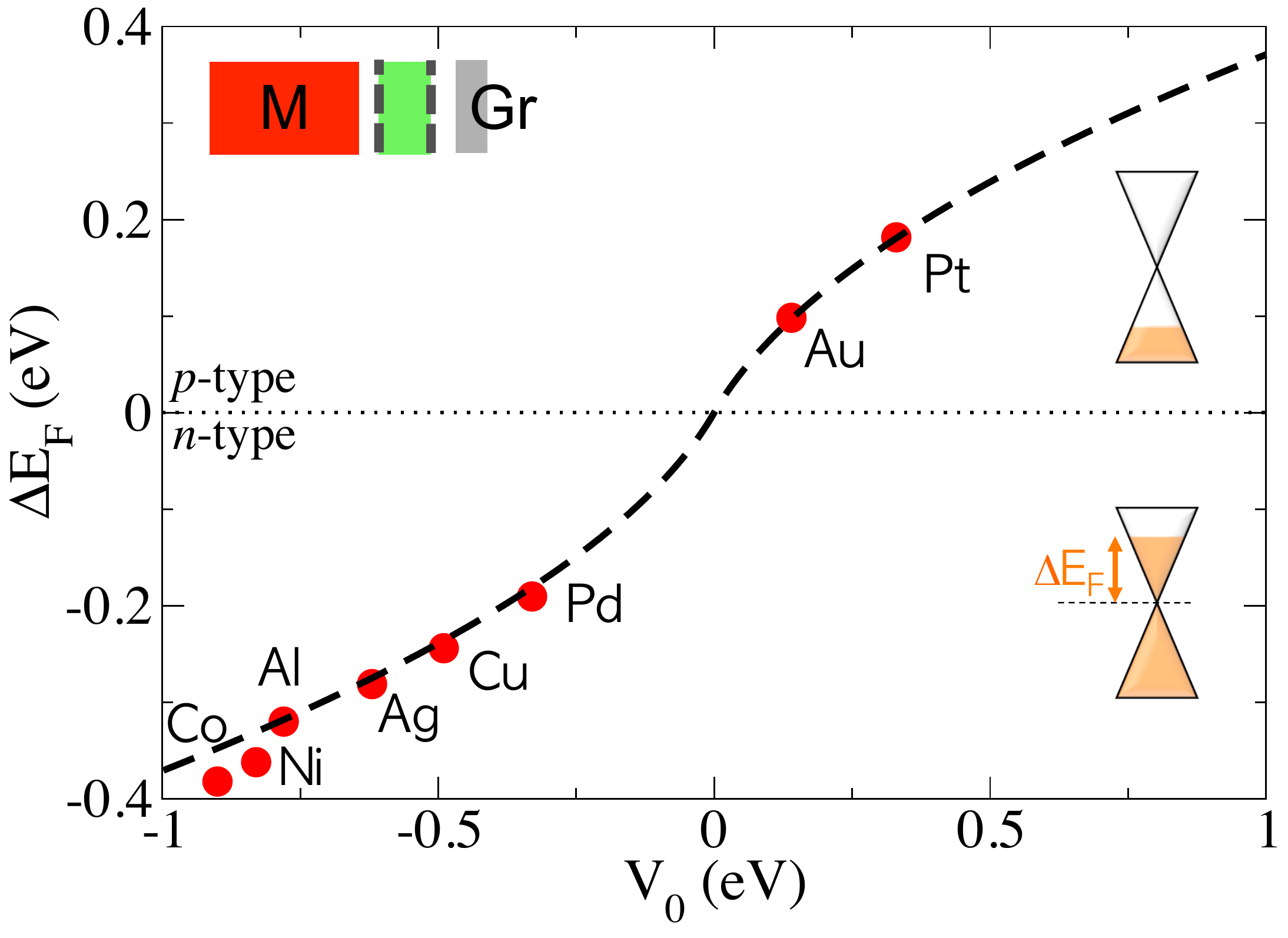}
\caption{(color online) The Fermi level shift for metal$|$\BN{}$|$graphene structures with 2 layers of \BN{}, plotted against $V_{\rm 0}$, according to  Eqs.~(\ref{modelV}) and (\ref{eq:V0}) with $V_\mathrm{g}=0$ and parameters from Table \ref{tbMBN}. The line is for a fixed $d$, and the circles are for the equilibrium metal$|$\BN{} bonding distances.} 
\label{fig:metals}
\end{figure}

For a dielectric consisting of two \BN{} monolayers, Figure \ref{fig:metals} shows how graphene in a metal$|$\BN{}$|$graphene structure is doped in the absence of an external field by different metals. Different metals have different workfunctions $W_{\rm M}$ and form different interface dipoles with the dielectric; each metal can be characterized by a different value of $V_0$ given by Eq.~(\ref{eq:V0}) using the parameters given in Table~\ref{tbMBN}. The dashed line represents the model of Eq.~(\ref{modelV}) for a fixed distance $d$, the circles correspond to the different metals where the equilibrium distances of Table~\ref{tbMBN} have been used. The predicted Fermi level shift $\Delta E_\mathrm{F}$ varies with the metal used for the gate electrode from +0.18 eV for Pt to $-$0.39 eV for Co. 

Most of the metals considered give rise to $n$-type doping of graphene; $p$-type doping is only found for the high work function metals Au and Pt. The amount of doping is however much less than what would be expected on the basis of the work function difference between these metals and graphene only. This is because of the large potential step $\Delta_\mathrm{M|BN}$ between Au or Pt and \BN{}, see Table~\ref{tbMBN}, which effectively lowers the metal work function. Although most of the other metals have a work function that is comparable to or higher than that of graphene, they result in $n$-type and not in $p$-type doping. Again this can be attributed to the effect of  the potential step $\Delta_\mathrm{M|BN}$. A similar effect was found for graphene that is directly adsorbed on a metal substrate.\cite{Giovannetti:prl08,Khomyakov:prb09} 

An external electric field can be used to counteract the intrinsic doping of graphene in these metal$|$\BN{}$|$graphene stacks and position the Fermi level at the charge neutrality point of graphene corresponding to $\Delta E_{\rm F}=0$. The field strengths required for the different metals cover a range of $\sim 0.5$ V/\AA. For the stacks considered in Fig.~\ref{fig:metals}, this corresponds to a range of 2 V in the gate voltage.

\section{Summary and conclusions}\label{sec:discussion}

We have developed a model, Eq.~(\ref{fermishiftE}), for the electrostatic doping of graphene in a field-effect device that describes the position of the Fermi level in graphene for a metal$|$dielectric$|$graphene structure as a function of the gate voltage. As input parameters the model needs the work function of the metal, the thickness and the dielectric constant of the dielectric material, and the potential steps formed at the metal$|$dielectric and the dielectric$|$graphene interfaces. All of these parameters can be obtained  from DFT calculations on single surfaces or interfaces, which makes the model parameter-free.

The model predicts doping of graphene even when the bias voltage is zero. This effect is caused by charge equilibration, i.e. the need to establish a common Fermi level across the metal$|$dielectric$|$graphene stack. The amount of doping is not only guided by the difference in work function between the metal and graphene, but also to a significant extent by the potential steps at the interfaces. These potential steps result from interface dipole layers that are formed by the chemical interactions at the interface. In particular the potential step at the metal$|$dielectric interface can be large, i.e., $>1$ eV, even when the bonding between the metal and the dielectric material is not strong. This potential step effectively lowers the work function of the metal, which causes metal substrates with work functions $\lesssim 5$ eV to dope graphene $n$-type. Only higher work function metal substrates lead to $p$-type doping. 

The amount of doping at zero bias decreases with increasing thickness of the dielectric. The Fermi level shift in graphene as a function of the applied electric field resulting from the gate voltage has a square-root like behavior. A similar behavior has been observed in experiments on gated $p$-Si$|$SiO$_2|$graphene structures, by  work function measurements,\cite{Yu:nanol09} and by conductivity measurements.\cite{Zhang:natp08} According to our model, this square-root like behavior is a direct result of the density of states in graphene being linear around the conical points and implies that the adsorption of graphene on the dielectric in those experiments was sufficiently weak as to leave the electronic structure of graphene essentially unchanged.

The model is tested by comparing its predictions to results obtained from explicit DFT calculations on metal$|$dielectric$|$graphene structures in the presence of an external electric field. We use the (111) or (0001) surfaces of the fcc or hcp close-packed metals Al, Co, Ni, Cu, Pd, Ag, Pt, Au, which span a range of work functions from 4.2 to 6.0 eV. As our dielectric we use either vacuum or \BN{}, which is an insulating material with a similar honeycomb structure as graphene, and we vary the thickness of the dielectric by varying the number of \BN{} layers. The model gives a good description of the Fermi level shift in graphene as a function of the applied electric field even for a \BN{} dielectric only a monolayer thick. The thickness dependence of the Fermi level shift is also reproduced very well by the model.

To be able to perform microscopic DFT calculations it was necessary to make approximations which are in general not satisfied by real materials. In particular, we assumed that graphene is commensurate with 
\BN{} and that both are commensurate with the underlying metal substrate. Where the in-plane bonding is very strong and the inter-plane bonding is weak as is the case for graphene on \BN{}, the resulting composite system remains incommensurate \cite{Xue:natm11,Decker:nanol11,Yankowitz:natp12,Joshi:nanol12} and it is not possible to carry out first-principles electronic structure calculations. Even when the bonding of graphene or \BN{} to a metal substrate is stronger, the periods that are found for commensurability are frequently so large \cite{Wintterlin:ss09} that DFT calculations are exorbitantly expensive. In such cases, we suggest that the parameters in the model we have outlined should be regarded as free parameters and determined by fitting to experimental observations. In this way, it should be possible to apply the model to very much more complex systems such as SiO$_2$ or SiC dielectrics and polycrystalline Si gates. Intrinsically planar dielectric spacers like MoS$_2$ and WS$_2$ that are currently of interest because of their potential in graphene electronics\cite{Britnell:sci12,Georgiou:arxiv12} are intermediate; they can be studied using DFT calculations, modelling the interfaces with metals and graphene using large lateral supercells, or the relevant parameters can be determined by fitting to experiment, or a combination of both.

\acknowledgements{The authors would like to thank Rien Wesselink, Zhe Yuan and Zhicheng Zhong for valuable discussions. MB and GB acknowledge support from the European project MINOTOR, grant no. FP7-NMP-228424. The use of supercomputer facilities was sponsored by the ``Stichting Nationale Computerfaciliteiten (NCF)'', financially supported by the ``Nederlandse Organisatie voor Wetenschappelijk Onderzoek (NWO)''.}

\appendix*

\section{Graphene Brillouin zone sampling}

The first-principles calculations presented in this paper were carried out with a $36 \times 36$ sampling of the two-dimensional (2D) Brillouin zone (BZ) for $(1 \times 1)$ graphene or an equivalent sampling density for $(2 \times 2)$ graphene. At energies close to the charge neutrality level of graphene, the density of states (DoS) is described well by the linear expression $D(E) = D_0|E|/A$ with $A$ the surface area of a graphene unit cell. Table~\ref{tab:D0} shows how the parameter $D_0$ depends on the number of $\mathbf{k}$-points used to sample the 2D-BZ. An unexpectedly dense grid ($288 \times 288$ points) is required to obtain a converged DoS which in view of the rather simple linear behavior of the DoS is at first glance surprising.  

\begin{table} [tb]
\begin{ruledtabular}
\begin{tabular}{rcccc}
      & \multicolumn{4}{c} {$D_0$ states/(eV$^2$ unit cell) } \\ \cline{2-5}
      & \multicolumn{2}{c} {Freestanding Graphene}  
                                    & \multicolumn{2}{c} {Graphene on $h$-BN}  \\ 
 $N$  &          $E<0$ & $E>0$                      & $E<0$   & $E>0$   \\ 
  \hline
  36  &          0.088 & 0.092                      &  0.102  & 0.102   \\
  72  &          0.112 & 0.119                      &  0.121  & 0.112   \\
 144  &          0.109 & 0.116                      &  0.114  & 0.115   \\
 288  &          0.108 & 0.115                      &  0.114  & 0.114   \\
\end{tabular} 
\end{ruledtabular}
\caption{Values of $D_0$ calculated using a grid obtained by dividing the reciprocal lattice vectors of the graphene lattice into $N$ intervals, and using the tetrahedron method\cite{Blochl:prb94a} to calculate the DoS. All grids contain the K-point.}
\label{tab:D0}
\end{table}

\begin{figure} [b]
\includegraphics[scale=0.14]{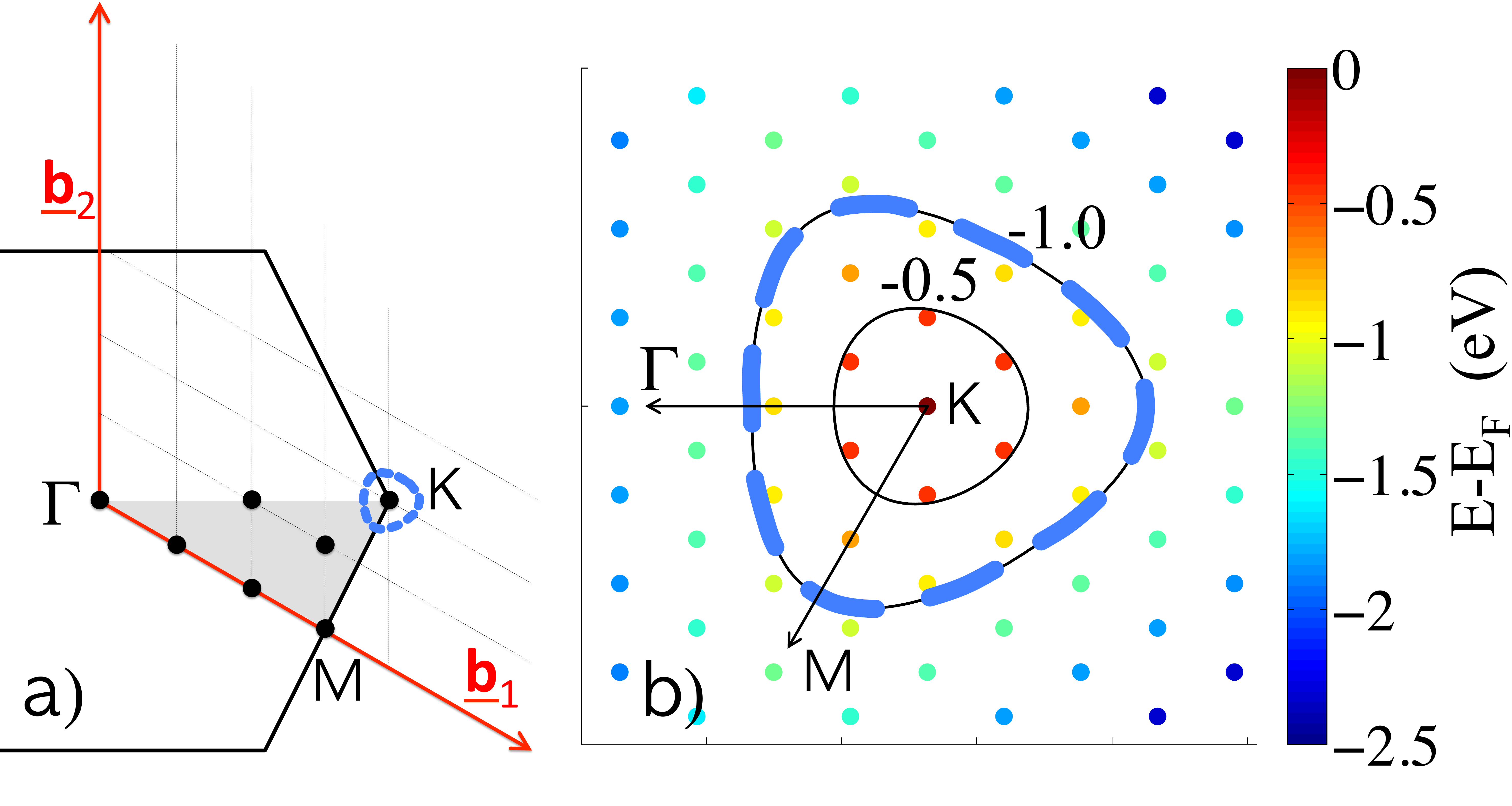}
\caption{(color online) a) Irreducible Brillouin zone (IBZ) of the graphene unit cell sampled with a $6\times6$ $\mathbf{k}$-point grid. b) The eigenvalues of free-standing graphene in the vicinity of the Dirac (K) point sampled with a $36 \times 36$ $\mathbf{k}$-point grid. The dashed contour about the K point in a) and b) represent a cut through the Dirac ``cone'' at -1.0~eV from the Dirac point while the solid contour in b) represents a cut at -0.5~eV.} \label{ibz}
\end{figure}

The main reason for requiring such a dense grid lies in the fact that for energies within 1 eV of the Fermi level, all of the states contributing to the DoS lie in a very small part of the BZ around the conical points. This is illustrated in Fig.~\ref{ibz} where it can be seen that even a $36 \times 36$ $\mathbf{k}$-point grid leads to a very coarse sampling of the relevant bands close to the conical points.

\begin{figure} [t]
\includegraphics[scale=0.33]{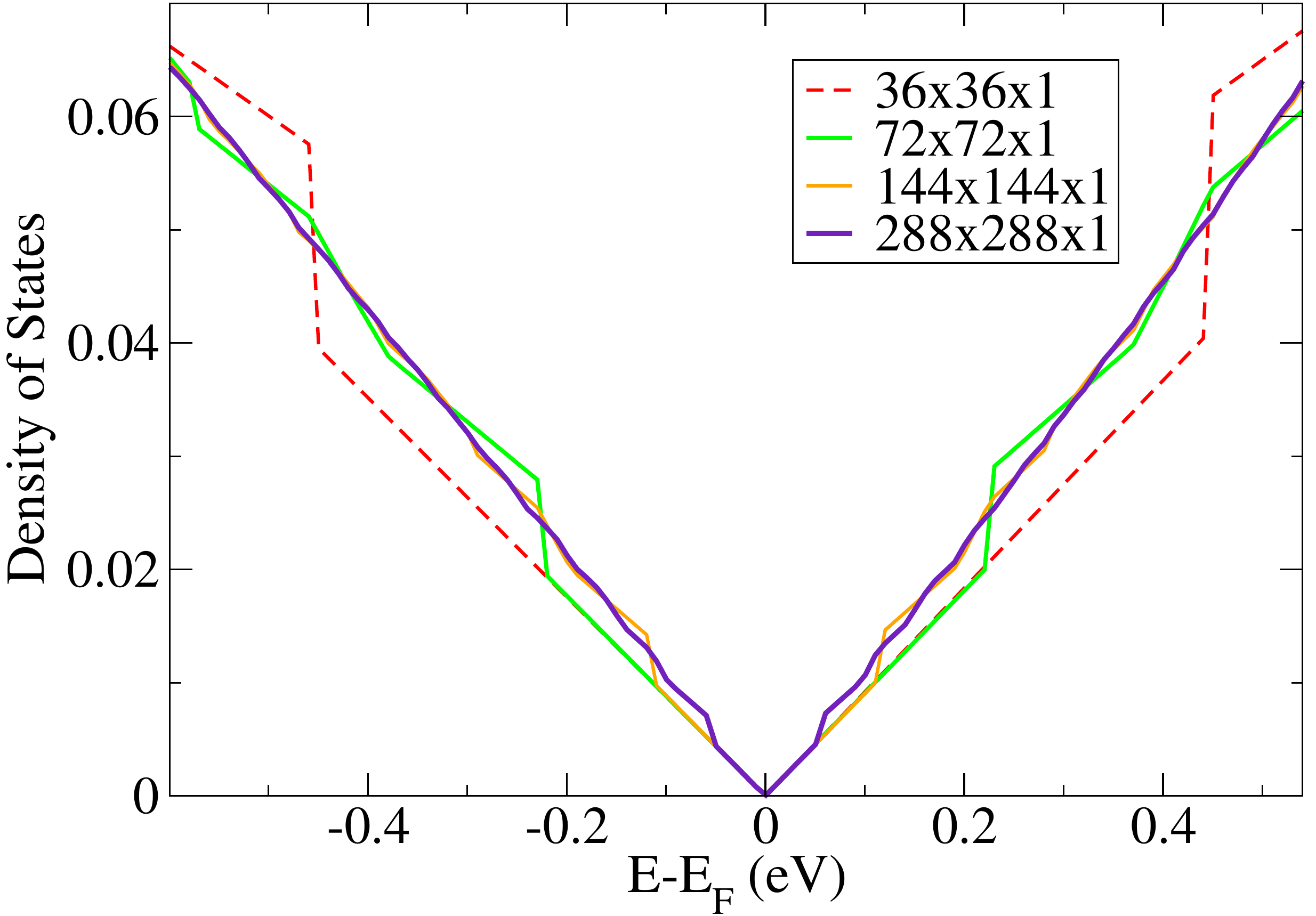}
\caption{(color online) DoS of graphene close to the Dirac point obtained by successively doubling the linear sampling density of the $\mathbf{k}$-point grid used in the tetrahedron scheme. With increasing grid density the DoS converges to a smooth line.} \label{dos}
\end{figure}

The DoS for various grid densities is shown in Figure \ref{dos}, calculated with the linear tetrahedron method\cite{Blochl:prb94a}, as implemented in VASP.\cite{Kresse:prb99,Kresse:prb93,Kresse:prb96}  For the $36\times36$ grid the DoS is perfectly linear up to $\pm 0.45$ eV, where it jumps to a different value. These jumps occur at eigenvalues corresponding to the $\mathbf{k}$-points that are explicitly included in the grid. With increasing grid density the number of jumps increase, but their size decreases, such that the DoS converges to a linear function. 

The jumps in the DoS at the $\mathbf{k}$-grid points are not an effect of the trigonal distortion of the graphene Dirac cone; the DoS of an ideal conical dispersion $E\propto|k|$ (centered on $\Gamma$) shows similar jumps when calculated with the tetrahedron method. The slopes $D_{\rm 0}$ listed in Table \ref{tab:D0} were extracted by fitting a linear curve to the DoS for energies $+0.4$ (or $-0.4$) eV with respect to the charge neutrality energy. For free-standing graphene the converged LDA values are 0.108 and 0.115 /(eV$^2$ unit cell) for the filled and empty states, respectively. 

\begin{figure} [b]
\includegraphics[scale=0.33]{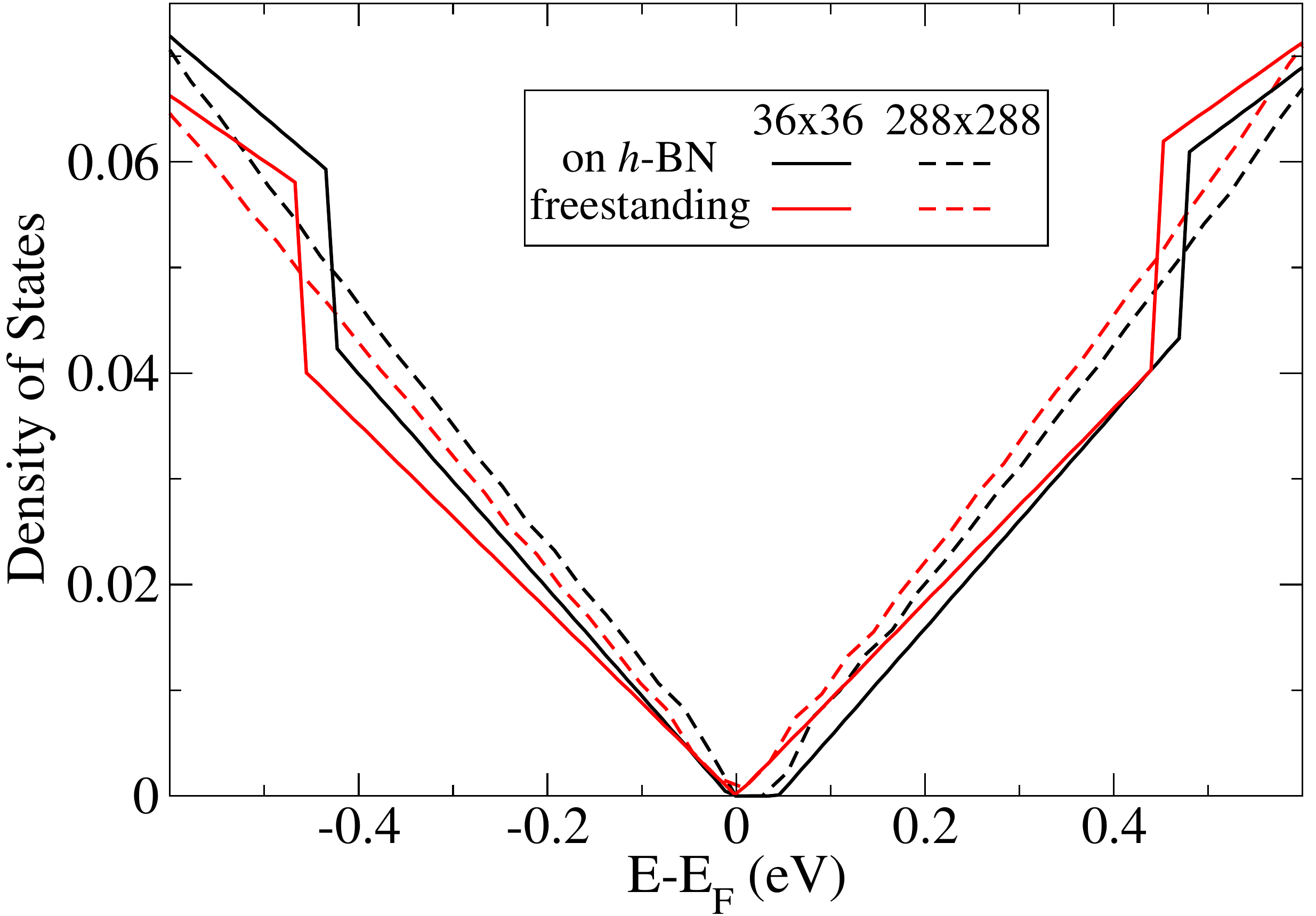}
\caption{(color online) DoS of free-standing graphene  (red/gray) and of graphene on \BN{} (black), close to the Dirac point. The solid and dashed curves were calculated using $36\times36$ and $288\times288$ grids, respectively.} \label{D0}
\end{figure}

\begin{figure} [h!]
\includegraphics[scale=0.4]{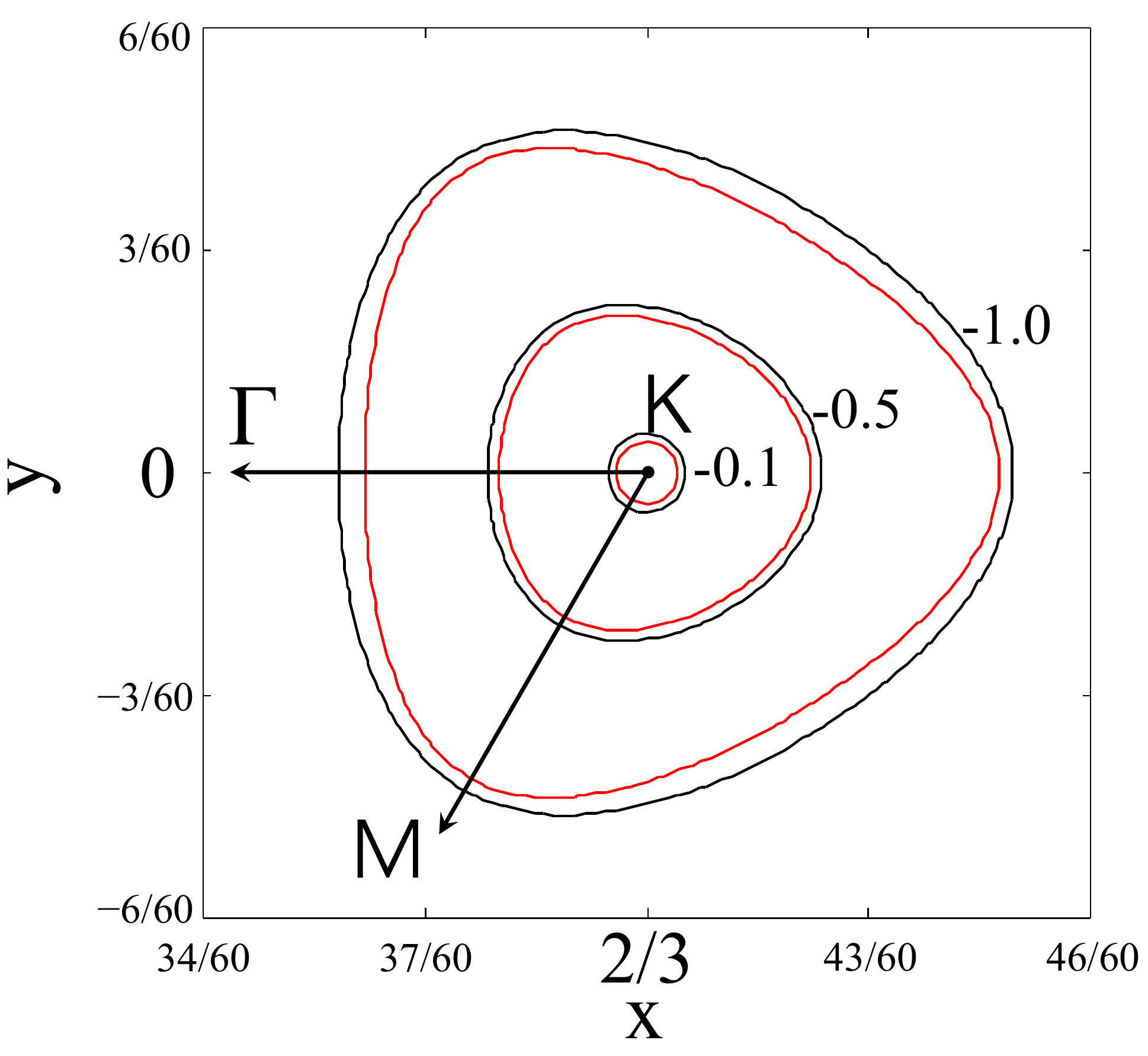}
\caption{(color online) Cross sections of the graphene Dirac cone at energies $-0.1,-0.5, -1.0$~eV with respect to the charge neutrality point, as calculated using a 288$\times$288 $\mathbf{k}$-point grid. The black lines correspond to graphene on \BN{} and the red(gray) lines to free-standing graphene. The axes are in units of $\nicefrac{2\pi}{a}$.} \label{conecross}
\end{figure}

The graphene DoS also depends somewhat on the substrate used, as illustrated by Fig.~\ref{D0} and Table \ref{tab:D0}. The slope $D_0$ for graphene adsorbed on \BN{} is somewhat larger than that of free-standing graphene for holes $E<0$ and somewhat smaller for electrons $E<0$. Curiously, the DoS is more symmetric for electrons and holes for graphene on \BN{}. A larger $D_0$ means that the graphene $\pi$-bands, and therefore the Dirac cone, become somewhat flattened, as illustrated in Fig. \ref{conecross}.

The DFT calculations on M$|$BN$|$Gr structures reported in this manuscript were carried out using a $36\times36$ $\mathbf{k}$-point grid. For internal consistency, a value $D_{0}=0.102$/(eV$^2$ unit cell) corresponding to this grid (see Table \ref{tab:D0}) was used in the model that fits the numerical results essentially perfectly.\cite{fn1}


\begin{thebibliography}{52}
\bibitem [{fn1()}]{fn1}%
  \BibitemOpen
  \href@noop {} {}\bibinfo {note} {The value we used for $D_0$ in
  Ref.~\onlinecite{Bokdam:nanol11}, $D_0=0.9$/(eV$^2$ unit cell), corresponds
  to the value calculated for freestanding graphene with a $36\times 36$
  $\mathbf{k}$-point grid, see Table \ref{tab:D0}. We use a slightly different
  value in this paper to be internally consistent with the explicit DFT
  calculations for $h$-BN performed with a $36\times 36$ $\mathbf{k}$-point
  grid.}\BibitemShut {Stop}%
\end{thebibliography}
\end{document}